\documentclass[prd,floatfix,onecolumn,amsmath,amssymb,10pt]{revtex4}
\usepackage{newtxtext,newtxmath} 
\usepackage{graphicx,color,dcolumn,booktabs,bm}
\usepackage{subfigure}
\usepackage{amssymb}
\usepackage{longtable}
\usepackage{indentfirst}
\usepackage{epsfig,gensymb,siunitx}
\usepackage{feynmf}   
\usepackage{epstopdf}   
\usepackage{slashed}  
\usepackage{cases}
\usepackage{xcolor}
\definecolor{navyblue}{RGB}{0,0,150}
\definecolor{CobaltBlue}{rgb}{0,0.28,.67}
\definecolor{maroon}{RGB}{139,25,150}
\usepackage{multirow}
\usepackage{float}
\usepackage{pgf}
\usepackage{graphicx,color,dcolumn,booktabs,bm}
\usepackage[colorlinks, citecolor=purple,anchorcolor=purple,menucolor=purple, linkcolor=purple,filecolor=purple,runcolor=purple,urlcolor=purple,frenchlinks=purple, urlcolor=purple]{hyperref}
\usepackage{orcidlink}
\usepackage{bbold}
\newcommand{\mtotal}{\mathcal{P}}

\newcommand{\ga}{\gamma_\alpha}
\newcommand{\gb}{\gamma_{\beta}}
\newcommand{\s}{\sigma_{\mu\nu}}

\begin{document}
	\preprint{}
	
	\title{\color{navyblue}{Tensor form factors of decuplet hyperons in QCD}}
	
	\author{Z.~Asmaee$^{1}$\orcidlink{0000-0002-3357-0574}}

	\author{K.~Azizi$^{1,2}$\orcidlink{0000-0003-3741-2167}}
	\email{kazem.azizi@ut.ac.ir}
	\thanks{Corresponding author}
	
	\affiliation{
		$^{1}$Department of Physics, \href{https://ut.ac.ir/en}{University of Tehran}, North Karegar Avenue, Tehran 14395-547, Iran\\
		$^{2}$Department of Physics, \href{https://www.dogus.edu.tr/en}{Dogus University}, Dudullu-\"{U}mraniye, 34775
		Istanbul,  T\"{u}rkiye}	
	\date{\today}
	
		\begin{abstract}
Tensor form factors encode essential information about the internal spin structure and tensor dynamics of baryons. In this work, we investigate the tensor form factors of the baryon hyperons $\Omega^-$, $\Sigma^{*+}$, and $\Xi^{*-}$ within the framework of QCD sum rules. The complete set of tensor form factors is numerically evaluated in the momentum transfer region $0<Q^2<10~\text{GeV}^2$. In addition, the quark tensor charges of the considered hyperons are extracted in the forward limit. The results provide new non-perturbative insight into the tensor structure and spin content of spin-$3/2$ baryons and offer valuable theoretical input for future phenomenological analyses and experimental studies.
	\end{abstract}
	
	\maketitle
	
		\section{Introduction}\label{Introduction} 
The internal structure of baryons continues to be a focus of both theoretical and experimental studies within the non-perturbative regime of quantum chromodynamics (QCD). In this context, form factors (FFs) play a central role, serving as fundamental inputs in the matrix elements of hadronic currents and providing a direct, quantitative connection between theoretical predictions and experimentally accessible observables.
Axial-vector FFs (AFFs) 
provide direct information on the helicity distributions of quarks and spin correlations inside baryons.
At low momentum transfer, electromagnetic FFs (EMFFs) are directly related to static electromagnetic (EM) properties such as magnetic moments and electric charge radii, while at higher momentum transfer they probe the spatial and dynamical distributions of quarks inside hadrons.
Gravitational FFs (GFFs), obtained from matrix elements of the QCD energy–momentum tensor (EMT), encode the mechanical properties of hadrons, such as the spatial distributions of mass, pressure, and shear forces.
Tensor FFs (TFFs), defined via matrix elements of the quark tensor current, provide complementary insight into the internal structure of hadrons. They are particularly sensitive to the transverse spin distributions of quarks and to spin–orbit correlations, revealing aspects of hadronic structure that are not accessible through other FFs. Together with EMFFs and GFFs; TFFs contribute to a complete description of hadron dynamics in the non-perturbative regime of QCD.
Over the past decades, extensive studies have been devoted to the investigation of FFs for hadronic systems with spin-$0$ \cite{Hohler:1974eq, Maris:2000sk, Broniowski:2008hx, Er:2022cxx, Esmer:2025xss}, $1/2$ \cite{Federbush:1958zz, Bincer:1959tz, Ma:2002ir, Aliev:2008cs, Aliev:2011ku, Azizi:2015fqa, Chen:2022odn, Dehghan:2025ncw, Najjar:2024deh, ShekariTousi:2024mso}, and $1$ \cite{Dong:2013rk, Sun:2017gtz, Polyakov:2019lbq, Sun:2020wfo}.
	
In this work, we focus on particles with spin-$3/2$,
which belong to the ten-dimensional representation commonly referred to as the decuplet baryons. This family includes the $\Delta$, $\Sigma^*$, $\Xi^*$, and $\Omega^-$ states.
The members of the decuplet baryons exhibit a wide range of masses and decay properties.
The $\Delta$ baryons are the lightest members of the decuplet baryons and decay almost exclusively via strong interactions into a nucleon and a pion. Their extremely short lifetimes \cite{ParticleDataGroup:2024cfk} make direct measurements of static properties, such as magnetic moments or charge radii, challenging.
Scattering experiments, including pion–nucleon \cite{LopezCastro:2000cv} and photon–nucleon \cite{Kotulla:2002cg} interactions, together with lattice QCD calculations \cite{Alexandrou:2008bn, Boinepalli:2009sq,Leinweber:1992hy,  Aubin:2008qp}, have provided valuable information on the properties of these baryons.
The $\Sigma^*$ and $\Xi^*$ baryons decay predominantly via strong interactions into a lower-mass baryon 
 and a pion or a kaon, resulting in very short lifetimes comparable to those of the $\Delta$ resonances.
The short lifetimes of these states pose significant challenges for direct experimental evaluation of static observables.
Consequently, most information about their EM properties comes from scattering experiments, including pion \cite{Seth:2014dxa, Junker:2019vvy, Dai:2021yqr}, as well as from lattice QCD \cite{Boinepalli:2009sq, Leinweber:1992hy,  Lin:2008rb}.

Experimental programs at facilities such as Jefferson Lab (JLab) have played an important role in exploring the EM structure of decuplet baryons. In particular, high-precision electron–nucleon scattering experiments have been extensively used to investigate the $\gamma^* N \to H$ transition FFs over a wide range of momentum transfers \cite{Bartel:1968tw, Stein:1975yy, Frolov:1998pw, Caia:2004pm, CLAS:2001cbm, CLAS:2003hro, CLAS:2009ces}, providing stringent constraints on theoretical models of the $\Delta$ resonance. For the strange members of the decuplet, including the $\Sigma^*$ and $\Xi^*$ baryons, experimental information has been obtained through kaon electroproduction and photoproduction processes \cite{Guo:2007dw, CLAS:2018kvn}, in which intermediate decuplet states contribute to the reaction mechanisms.
The $\Omega^-$ baryon, unlike the other decuplet states, decays only via the weak interaction, giving it a significantly longer lifetime than the $\Delta$, $\Sigma^*$, or $\Xi^*$ baryons. This longer lifetime allows for more direct experimental investigations of its internal structure, including EM and static properties, which have been probed in the process $e^- e^+ \to \Omega^- \bar{\Omega}^+$ by the CLEO \cite{Dobbs:2014ifa} and BESIII \cite{BESIII:2022kzc} collaborations, alongside lattice QCD calculations \cite{Leinweber:1992hy, Aubin:2008qp, Alexandrou:2010jv}.	
Both experimental groups, such as the A1 and A2 Collaborations \cite{Beck:1997ew, Pascalutsa:2005ts, A1:2008ocu, Sparveris:2013ena} and the LEGS Collaboration \cite{Blanpied:1997zz, Blanpied:2001ae}, as well as a variety of theoretical approaches, including QCD sum rules (QCDSR) \cite{Lee:1997jk, Aliev:2009pd, Azizi:2009egn, Aliev:2015qea}, $1/N_c$ expansion \cite{Buchmann:2002et, Flores-Mendieta:2015wir, BandaGuzman:2020fxx, GonzalezdeUrreta:2012zz, Matagne:2012zz, Goity:2010zz, Jayalath:2010zz}, the chiral quark soliton and constituent models \cite{Kim:2019gka, Geng:2009ys, Berger:2004yi}, the Skyrme and quark models \cite{Oh:1995hn, Krivoruchenko:1991pm, Wang:2023bjp}, have also investigated the EM properties of decuplet baryons.

GFFs characterize the interaction between matter fields and gravity, where scattering off a graviton would be a natural but experimentally inaccessible probe. However, because the EMT has a structure similar to the EM current \cite{Diehl:2003ny}, hard-exclusive processes provide a practical way to access hadron GFFs via generalized parton distributions (GPDs) and generalized distribution amplitudes \cite{Polyakov:2019lbq, Kumano:2017lhr}.	
The inherently short life of decuplet baryons makes the extraction of GPDs experimentally challenging, leaving their GFFs far less studied than their EMFFs.
Various non-perturbative approaches have been applied to investigate the GFFs and mechanical properties of decuplet baryons \cite{ Wang:2023bjp, Fu:2022rkn, Kim:2020lrs, Pefkou:2021fni, Panteleeva:2020ejw, Dehghan:2023ytx, Dehghan:2025eov, Perevalova:2016dln, Cotogno:2019vjb, Panteleeva:2020ejw, Pefkou:2021fni, Kim:2021zbz, Fu:2022rkn, Alharazin:2022wjj, Alharazin:2022xvp, Fu:2023ijy, Kim:2022bwn, Ozdem:2022zig, Kim:2023yhp, Alharazin:2023zzc, Goharipour:2024atx}.
	
At leading twist, hadron structure is encoded in three independent quark distributions, each probing a distinct aspect of quark dynamics. The  unpolarized distribution describes the momentum distribution of quarks inside the hadron without reference to spin, while the helicity distribution quantifies the longitudinal spin alignment of quarks in polarized hadrons. Owing to their chiral-even character, both distributions are readily accessible through inclusive deep inelastic scattering. The third distribution, known as transversity, provides insight into the transverse spin structure of hadrons by encoding the correlation between the transverse polarization of quarks and that of the parent hadron. Because transversity is chiral-odd and related to the tensor current, it decouples from inclusive measurements and can only be probed through processes involving another chiral-odd quantity. Consequently, its experimental determination relies on more exclusive reactions, such as semi-inclusive deep inelastic scattering and Drell–Yan processes, where transversity enters at leading twist.	
The tensor operator plays a fundamental role in the study of hadron structure, as its matrix elements are directly related to the transversity distribution, which probes the transverse spin degrees of freedom of quarks \cite{Ralston:1979ys, Jaffe:1991kp, Jaffe:1991ra, Barone:2001sp}.	
It has also been established that hadronic TFFs can be interpreted as generalized FFs, obtained from the Mellin moments of GPDs. Within this framework, TFFs provide direct access to the transverse spin structure of quarks in hadrons, offering a probabilistic description of quark transverse densities in the impact-parameter space.
In particular, at zero momentum transfer, the lowest Mellin moment of the chiral-odd GPDs reduces to the tensor charge, which has been investigated theoretically using various approaches,  including the bag model \cite{He:1994gz} and quark models \cite{Pasquini:2005dk, Gamberg:2001qc}.
TFFs play a crucial role in advancing our understanding of baryon tomography, as they provide essential information on the transverse spin structure, spin–orbit correlations, and relativistic effects inside baryons. Consequently, the study of TFFs is essential for achieving a more complete understanding of hadron structure, providing insights beyond those accessible through other FFs.
	
Investigations of TFFs to date have focused on octet hyperons \cite{kucukarslan:2016xhx} and nucleons \cite{Aliev:2011ku, Gutsche:2016xff, Azizi:2019ytx, Ozdem:2020vpt, Ozdem:2021zbn}.
While nucleons have been extensively studied both theoretically and experimentally, investigations of TFFs for decuplet baryons remain relatively limited.	
In this context, the TFFs of decuplet baryons have recently received attention, in Ref. \cite{Fu:2024kfx, Fu:2025upc} presenting a decomposition based on the first Mellin moments of their transversity GPDs.
In our previous work \cite{Asmaee:2025elo}, we presented the tensor current matrix elements using a method entirely independent of transversity GPDs and investigated the corresponding TFFs of the $\Delta^+$ baryon.
In this study, we investigate the TFFs of the $\Omega^-$, $\Sigma^{*+}$, and $\Xi^{*-}$ baryons within the framework of the QCDSR method. The organization of the paper is as follows: Sec. \ref{QCDSR} is devoted to the calculation of the TFFs of the baryon hyperons $\Omega^-$, $\Sigma^{*+}$, and $\Xi^{*-}$ based on the three-point correlation function.
The physical and QCD sides of the correlation function are discussed in Sections.~\ref{physical side} and \ref{QCD side}, respectively.
In Sec.~\ref{QCD side}, we study the correlation function for the baryon hyperons.
In Sec.~\ref{ANALYSES}, the coefficients of the relevant Lorentz structures are matched between the physical and QCD sides. We then perform a numerical analysis of the momentum transfer dependence of the TFFs for the $\Omega^-$, $\Sigma^{*+}$, and $\Xi^{*-}$ baryons and extract the corresponding tensor charges.
	
\section{TFFs of the baryon hyperons in QCDSR}\label{QCDSR}	
To study the TFFs of the considered baryon hyperons $\Omega^-$, $\Sigma^{*+}$, and $\Xi^{*-}$ within the framework of the QCDSR approach, we evaluate the three-point correlation function induced by the tensor current \cite{Dehghan:2023ytx, Dehghan:2025eov, Asmaee:2025elo},	
\begin{equation}
	\Pi^H_{\alpha\mu\nu\beta}(p,p') = i^2 \int d^4 x\ e^{-ip.x} \int d^4 y\ e^{ip'.y}
	\langle 0 |\mathcal{T}[J_{\alpha}^{H}(y)J^{T,\ H}_{\mu \nu}(0)\bar{J}_{\beta}^{H}(x)]| 0 \rangle,
	\label{corrf}
\end{equation}	
where, $\mathcal{T}$ stands for the time-ordering operator, and $J^{T,\ H}_{\mu\nu}$ (with $H=\Omega^-, \Sigma^{*+}, \Xi^{*-}$) denotes the tensor current of the baryon hyperon under consideration.
The momenta of the initial and final baryon states are given by $p$ and $p'$, respectively.
The interpolating current $J^{H}_\alpha(y)$ associated with the baryon hyperons at position $y$ and carrying Lorentz index $\alpha$, is defined as \cite{Aliev:2016jnp, Aliev:2010dw, Azizi:2016ddw},
\begin{equation}
	\begin{split}
		&J_{\alpha}^{H}(y)=A\ \varepsilon^{abc}\Bigg[
		 \Big(\psi_1^{aT} (y) C\gamma_{\alpha}\psi_2^{b}(y) \Big)  \psi_3^{c}(y)
		 +\Big(\psi_2^{aT}(y) C\gamma_{\alpha}\psi_3^{b} (y) \Big)\psi_1^{c}(y)
		 +\Big(\psi_3^{aT} (y) C\gamma_{\alpha}\psi_1^{b} (y) \Big)  \psi_2^{c}(y)\Bigg],
	\end{split}
	\label{J-ab}
\end{equation}	
where $C$ is the charge-conjugation operator, and $a$, $b$, and $c$ denote the color indices.
The normalization constant $A$ and the light-quark fields $\psi_1$, $\psi_2$, and $\psi_3$ \cite{Aliev:2010dw} are listed in Table \ref{hyperon}.
The tensor current $J^{T,\ H}_{\mu\nu}(0)=\bar{\psi}(0)i\sigma_{\mu\nu} \psi(0)$ relevant for baryon hyperons is given by \cite{kucukarslan:2016xhx, Aliev:2011ku,  Asmaee:2025elo},
	\allowdisplaybreaks
	\begin{align}
		&J^{T,\ \Omega^-}_{\mu\nu}(0)=\bar{s}(0)i\sigma_{\mu\nu} s(0),\nonumber
		\\
		&J^{T,\ \Sigma^{*+}}_{\mu\nu}(0)=\bar{u}(0)i\sigma_{\mu\nu} u(0)+\bar{s}(0)i\sigma_{\mu\nu} s(0),
	\nonumber	\\
		&J^{T,\ \Xi^{*-}}_{\mu\nu}(0)=\bar{s}(0)i\sigma_{\mu\nu} s(0)+\bar{d}(0)i\sigma_{\mu\nu} d(0),
		\label{tensor-current}
	\end{align}
where, $\sigma_{\mu\nu}=\frac{i}{2}\big[\gamma_\mu,\gamma_\nu\big]$ represents the antisymmetric tensor constructed from the gamma matrices.

\begin{table}[!htb]
	\centering
	\begin{minipage}{0.60\textwidth} 
		\centering
		\renewcommand{\arraystretch}{1.4} 
		\setlength{\tabcolsep}{12pt}      
		\begin{tabular}{|c|c|c|c|c|}
			\hline
			\text{baryon hyperons} & $A$ & $\psi_1$ & $\psi_2$ & $\psi_3$ \\
			\hline\hline
			$\Omega^-$ & $1/{3}$ & $s$ & $s$ & $s$ \\
			\hline
			$\Sigma^{*+}$ & $1/{\sqrt{3}}$ & $u$ & $u$ & $s$ \\
			\hline
			$\Xi^{*-}$ & $1/{\sqrt{3}}$ & $s$ & $s$ & $d$ \\
			\hline
		\end{tabular}
	\end{minipage}
	\caption{Values of the normalization constants $A$ and the corresponding light-quark fields $\psi_1$, $\psi_2$, and $\psi_3$ for the baryon hyperons studied in this work.}
	\label{hyperon}
\end{table}
The TFFs of baryon hyperons are investigated within the QCDSR framework by evaluating the correlation function in Eq.~\eqref{corrf} in two equivalent but distinct representations.
In the hadronic representation, the correlator is parameterized in terms of physical observables associated with the baryon states, while in the QCD representation it is expressed through quark and gluon dynamics.
In order to suppress higher states and continuum contributions while enhancing the ground state baryon hyperons, double Borel transformations with respect to the squared momenta $p^2$ and $p^{\prime 2}$ are performed on both sides, together with continuum subtraction.
The TFFs are then obtained by matching the coefficients of the same Lorentz structures appearing in both representations.
	
\subsection{Physical side}\label{physical side}
Within the QCDSR framework, the physical (hadronic) side of the correlation function is evaluated by inserting a complete set of intermediate states whose quantum numbers match those of the interpolating current for the system under study \cite{Dehghan:2023ytx, Dehghan:2025eov, Asmaee:2025elo, Najjar:2025dzl},
\begin{equation}
	\begin{split}
		&1 = \left| 0\left\rangle \right\langle 0\right| +\sum_l \int \frac{d^4p_l}{(2\pi)^4} (2\pi) \delta(p_l^2 - m^2_H) \left| H(p_l)\rangle \langle H(p_l)\right| + \text{higher Fock states},
		\label{set complet}
	\end{split}
\end{equation}	
where $m_H$ represents the masses of the baryon hyperons, and $|H(p_l)\rangle$ denotes the hadronic state of interest, carrying four-momentum $p_l$.	
For the initial and final baryon hyperons, carrying four-momenta $p_l$ and $p'_l$, Eq.~\eqref{set complet} is applied separately, and the resulting expression is integrated over the four-dimensional space-time coordinates $x$ and $y$ using the identity \cite{Dehghan:2025eov},	
\begin{equation}
	\begin{split}
	\int d^4x \int \frac{d^4p_l}{(2\pi)^4} (2\pi) \delta(p_l^2 - m^2_H) e^{i(p_l-p).x}=\dfrac{i}{m_H^2-p^2}.
	\end{split}
\end{equation}
This procedure isolates the ground state pole contributions for both the initial and final baryon hyperons.	
Upon completing the necessary calculations, the correlation function of Eq. \eqref{corrf} on the physical side for the considered baryon hyperons  is obtained in the form \cite{Dehghan:2023ytx, Dehghan:2025eov, Asmaee:2025elo}:
\begin{equation}
	\Pi_{\alpha\mu\nu\beta}^{\text{phy},\ H}(p,p^\prime) = 
	\sum_{s,s{'}}\frac{\langle0|J_{\alpha}^{H}(0)|{H(p',s')}\rangle\langle {H(p',s')}
		|J^{T,\ H}_{\mu \nu}(0)|
		{H(p,s)}\rangle\langle {H(p,s)}
		|\bar{J}_{\beta}^{H}(0)| 0 \rangle}
	{(m^2_{H}-p^2)(m^2_{H}-p'^2)} 
	+\cdots,
	\label{physicalside correlation function}
\end{equation}	
where, $s$ and $s^\prime$ indicate the spin of the initial and final states, respectively.	
The baryon hyperon states are described by $\left| H(p,s)\right\rangle $ for the initial state and $\left| H(p',s')\right\rangle $ for the final state.
The dots denote contributions arising from excited states and the continuum.
The first and third matrix elements of the interpolating current between the vacuum and the baryon hyperon appearing in Eq.~\eqref{physicalside correlation function} are defined as \cite{Aliev:2011ku, Aliev:2016jnp, Ioffe:1981kw, Aliev:2002ra, Azizi:2014yea},	
\begin{equation}
	\begin{split}
		&\langle0|J_{\alpha}^{H}(0)|{H(p',s')}\rangle=\lambda_H u_{\alpha}(p^\prime, s^\prime),
		\label{residue}
	\end{split}
\end{equation}
where $\lambda_H$ denotes the residue of the hyperon baryon, and $u_\alpha(p',s')$ is the Rarita–Schwinger spinor. The sum over its spin states is given by \cite{Dehghan:2023ytx, Dehghan:2025eov, Asmaee:2025elo, Aliev:2016jnp}.
	\begin{equation}
	\begin{split}
		&	\sum_{s'}{u_\alpha}(p',s') \bar{u}_{\alpha'}(p',s') = 
		- ({\slashed{p}^\prime} + m_H)
		\Big[g_{\alpha\alpha'} - \frac{\gamma_{\alpha}\gamma_{\alpha'}}{3}
		-\frac{2 p'_{\alpha} p'_{\alpha'}}{3 m_H^2}
		+\frac{p'_{\alpha} \gamma_{\alpha'} - p'_{\alpha'} \gamma_{\alpha}}{3 m_H}
		\Big].
	\end{split}
	\label{sum spin}
\end{equation}
The second matrix element in Eq.~\eqref{physicalside correlation function} represents the tensor current between baryon hyperon states and is expressed in terms of the corresponding TFFs as \cite{Asmaee:2025elo},	
\allowdisplaybreaks
\begin{align}
	\left\langle H(p^\prime,s^\prime)\left| J_{\mu\nu}^{T,\ H}(0)\right| H(p,s)\right\rangle&=\bar{u}^\alpha(p^\prime,s^\prime)
\Bigg[i\sigma_{\mu\nu} \bigg( g_{\alpha\beta}F^{T,\ H}_{1,0}(Q^2)
+\frac{q_{\alpha}q_{\beta}}{m_H^2}F^{T,\ H}_{1,1}(Q^2)\bigg) 
+ig_{\alpha [\mu}\sigma _{\nu]\beta}F_{2,0}^{T,\ H}(Q^2)\nonumber\\
&+\frac{\gamma_{[\mu}q_{\nu]}}{m_H}\bigg(g_{\alpha\beta}F_{3,0}^{T,\ H}(Q^2)
+\frac{q_{\alpha}q_{\beta}}{m_H^2}F^{T,\ H}_{3,1}(Q^2) \bigg) 
+\frac{\gamma_{[\mu}\mtotal_{\nu]}\mtotal_{[\alpha}q_{\beta]}}{m_H^3} F^{T,\ H}_{4,0}(Q^2)\nonumber\\
&+\frac{\mtotal_{[\mu}q_{\nu]}}{m_H^2}\bigg(g_{\alpha\beta}F_{5,0}^{T,\ H}(Q^2)
+\frac{q_{\alpha}q_{\beta}}{m_H^2}F^{T,\ H}_{5,1}(Q^2) \bigg)
+\frac{g_{\alpha[\mu}\gamma_{\nu]}q_\beta-g_{\beta[\mu}\gamma_{\nu]}q_\alpha}{m_H}F^{T,\ H}_{6,0}(Q^2)\nonumber\\
&+\frac{g_{\alpha[\mu}q_{\nu]}\mtotal_\beta-g_{\beta[\mu}q_{\nu]}\mtotal_\alpha}{m_H^2}F^{T,\ H}_{7,0}(Q^2)
\Bigg]u^\beta(p,s),
\label{martix-element}
\end{align}	
where $g_{\alpha\beta}$ is the metric tensor, the momentum transfer and total momentum are defined as
$q_\mu=p_\mu-p^\prime_\mu$ and $\mtotal_\mu=p_\mu+p^\prime_\mu$, 
respectively, with $Q^2=-q^2$. 
The tensor form factors $F^{T,\ H}_{i,j}(Q^2)$ correspond to the considered baryon hyperons, and the antisymmetric combination of indices is defined as $a_{[\mu}b_{\nu]}=a_\mu b_\nu-a_\nu b_\mu$.
By inserting Eqs.~\eqref{residue}, \eqref{sum spin}, and \eqref{martix-element} into Eq.~\eqref{physicalside correlation function}, the physical side of the correlation function for the hyperon–hyperon transition is obtained.	
While the interpolating current primarily couples to the spin-$3/2$ baryon hyperon, its Lorentz structure also permits overlap with spin-$1/2$ states, giving rise to unwanted contributions, that can be expressed as \cite{Dehghan:2023ytx, Dehghan:2025eov, Asmaee:2025elo, Aliev:2016jnp},	
\allowdisplaybreaks
\begin{align}
&	\langle0\mid J^H_{\alpha}(0)\mid p', s'=1/2\rangle=(A  p^\prime_{\alpha}+B\gamma_{\alpha})u(p^\prime,s^\prime=1/2),
\label{unwnted}
\end{align}	
where, $u(p', s' = 1/2)$ denotes the Dirac spinor, while $A$ and $B$ correspond to scalar functions.	
As seen in Eq.~\eqref{unwnted}, the spin-$1/2$ contamination, proportional to $p'_\alpha$ and $\gamma_\alpha$, is eliminated by imposing the constraints $p^{\prime \alpha} J_\alpha^H = 0$ and $\gamma^\alpha J_\alpha^H = 0$.
In this way, $A$ can be written in terms of $B$ $\left(\text{\textit{i.e};}\ A=-4B/m_{1/2} \right) $, which removes the spin-$1/2$ contamination and isolates the pure spin-$3/2$ contributions in the correlation function.	
In practical calculations, the Dirac matrices are first arranged in the order $\gamma_{\alpha} \slashed{p}' \slashed{p} \gamma_{\mu} \gamma_{\nu} \gamma_{\beta}$, after which all terms beginning with $\gamma_\alpha$, ending with $\gamma_\beta$, or proportional to $p'_\alpha$ and $p_\beta$ are eliminated.	
Finally, the effects of excited states and the continuum are reduced by applying a double Borel transformation respect to $p^2$ and $p^{\prime 2}$, with Borel parameters $M_1^2$ and $M_2^2$ \cite{Ozdem:2017jqh, Azizi:2018duk},	
\begin{equation}
	\mathcal{B}_{M_1^2}	\mathcal{B}_{M_2^2}\left( \dfrac{1}{\left( m_H^2-p^{2}\right) \left( m_H^2-p^{\prime 2}\right) }\right) =e^{-\frac{m_H^2}{M_1^2}}e^{-\frac{m_H ^2}{M_2^2}}=e^{-\frac{m_H^2}{M^2}}.
	\label{Borel}
\end{equation}
Because both the initial and final states involve the same baryon hyperon, we set the Borel masses equal, $M_1^2 = M_2^2 = 2M^2$, as used in the second equality of Eq.~\eqref{Borel}.
Upon completing the above steps, the physical side of the correlation function for the baryon hyperons $\Omega^-$, $\Sigma^{*+}$, and $\Xi^{*-}$ takes its final form in the Borel scheme, expressed as follows:
\allowdisplaybreaks
\begin{align}
		\Pi_{\alpha\mu\nu\beta}^{\text{phy},\ \Omega^-}(M^2,Q^2) = \left|\lambda_{\Omega^-} \right|^2 e^{-\frac{m_{\Omega^-}^2}{M^2}}&\bigg[ g_{\alpha\beta}\gamma_{\mu}\gamma_\nu\slashed{p}^\prime \slashed{p}\Pi_1^{\text{phy},\ \Omega^-}(Q^2)
	+g_{\alpha\mu}g_{\nu\beta}\slashed{p}^\prime\Pi_2^{\text{phy},\ \Omega^-}(Q^2)
	+ p_\alpha  \gamma_{\mu}\gamma_\nu p^\prime_\beta\slashed{p}^\prime \slashed{p}\Pi_3^{\text{phy},\ \Omega^-}(Q^2)\nonumber\\
	&
	+p'_\mu g_{\alpha\beta}p_\nu\slashed{p}\Pi_4^{\text{phy},\ \Omega^-}(Q^2)
	+p_\alpha g_{\mu\beta}\gamma_\nu \slashed{p}^\prime\Pi_5^{\text{phy},\ \Omega^-}(Q^2)
	+p_\alpha p^\prime_\mu g_{\nu\beta} \slashed{p}^\prime\Pi_6^{\text{phy},\ \Omega^-}(Q^2)\nonumber\\&
	+p_\alpha p_\mu p^\prime_\nu p^\prime_\beta \slashed{p}^\prime\Pi_{7}^{\text{phy},\ \Omega^-}(Q^2)
	+p_{\mu}g_{\alpha\beta}p^\prime_\nu \mathbb{1}\Pi_{8}^{\text{phy},\ \Omega^-}(Q^2)
	+p_\alpha p_\mu \gamma_\nu p^\prime_\beta \slashed{p} \Pi_{9}^{\text{phy},\ \Omega^-}(Q^2)\nonumber\\
	&
	+p_\alpha p^\prime_\mu \gamma_\nu p^\prime_\beta \slashed{p}\Pi_{10}^{\text{phy},\ \Omega^-}(Q^2)
	+\cdots\bigg],
	\label{phy omega}
\end{align}	
\allowdisplaybreaks
\begin{align}
	\Pi_{\alpha\mu\nu\beta}^{\text{phy},\ \Sigma^{*+}}(M^2,Q^2) = \left|\lambda_{\Sigma^{*+}} \right|^2 e^{-\frac{m_{\Sigma^{*+}}^2}{M^2}}&\bigg[g_{\alpha\beta}\gamma_{\mu}\gamma_\nu\slashed{p}^\prime \slashed{p}\Pi_1^{\text{phy},\ \Sigma^{*+}}(Q^2)
	+g_{\alpha\mu}g_{\nu\beta}\slashed{p}^\prime \slashed{p}\Pi_2^{\text{phy},\ \Sigma^{*+}}(Q^2)
	+ p_\alpha \gamma_{\mu}\gamma_\nu p^\prime_\beta\slashed{p}^\prime \slashed{p}\Pi_3^{\text{phy},\ \Sigma^{*+}}(Q^2)
	\nonumber\\
	&+g_{\alpha\beta} \gamma_\mu p^\prime_\nu\slashed{p}^\prime \slashed{p}\Pi_4^{\text{phy},\ \Sigma^{*+}}(Q^2)
	+g_{\alpha\beta} p_\mu p^\prime_\nu \slashed{p}^\prime\slashed{p}\Pi_5^{\text{phy},\ \Sigma^{*+}}(Q^2)
	+p_\alpha g_{\mu\beta}\gamma_\nu \slashed{p}^\prime\slashed{p}\Pi_6^{\text{phy},\ \Sigma^{*+}}(Q^2)
	\nonumber\\
	&+p_\alpha p_\mu g_{\nu\beta} \slashed{p}^\prime\slashed{p}\Pi_7^{\text{phy},\ \Sigma^{*+}}(Q^2)
	+p_\alpha p'_\mu p_\nu p'_\beta  \mathbb{1} \Pi_8^{\text{phy},\ \Sigma^{*+}}(Q^2)
	+p_\alpha p_\mu\gamma_\nu p^\prime_\beta \slashed{p}\Pi_{9}^{\text{phy},\ \Sigma^{*+}}(Q^2)
	\nonumber\\
	&
	+p_\alpha p'_\mu\gamma_\nu p^\prime_\beta \slashed{p}\Pi_{10}^{\text{phy},\ \Sigma^{*+}}(Q^2)
	+\cdots\bigg],
	\label{phy sigma}
\end{align}
\allowdisplaybreaks
\begin{align}
\Pi_{\alpha\mu\nu\beta}^{\text{phy},\ \Xi^{*-}}(M^2,Q^2) = \left|\lambda_{\Xi^{*-}} \right|^2 e^{-\frac{m_{\Xi^{*-}}^2}{M^2}}&\bigg[
g_{\alpha\beta}\gamma_{\mu}\gamma_\nu\slashed{p}^\prime \slashed{p}\Pi_1^{\text{phy},\ \Xi^{*-}}(Q^2)
+g_{\alpha\mu}g_{\nu\beta}\slashed{p}^\prime \slashed{p}\Pi_2^{\text{phy},\ \Xi^{*-}}(Q^2)
+ p_\alpha \gamma_{\mu}\gamma_\nu p^\prime_\beta\slashed{p}^\prime \slashed{p}\Pi_3^{\text{phy},\ \Xi^{*-}}(Q^2)
\nonumber\\
&+g_{\alpha\beta} \gamma_\mu p^\prime_\nu\slashed{p}^\prime \slashed{p}\Pi_4^{\text{phy},\ \Xi^{*-}}(Q^2)
+g_{\alpha\beta} p_\mu p^\prime_\nu \slashed{p}^\prime\slashed{p}\Pi_5^{\text{phy},\ \Xi^{*-}}(Q^2)
+p_\alpha g_{\mu\beta}\gamma_\nu \slashed{p}^\prime\slashed{p}\Pi_6^{\text{phy},\ \Xi^{*-}}(Q^2)\nonumber\\
&
+p_\alpha p'_\mu g_{\nu\beta} \slashed{p}^\prime\Pi_7^{\text{phy},\ \Xi^{*-}}(Q^2)
+p_\alpha p'_\mu p_\nu p'_\beta  \mathbb{1}\Pi_{8}^{\text{phy},\ \Xi^{*-}}(Q^2)
+p_\alpha p_\mu\gamma_\nu p^\prime_\beta \slashed{p}\Pi_{9}^{\text{phy},\ \Xi^{*-}}(Q^2)
\nonumber\\
&+p_\alpha p'_\mu\gamma_\nu p^\prime_\beta \slashed{p}\Pi_{10}^{\text{phy},\ \Xi^{*-}}(Q^2)
+\cdots\bigg],
\label{phy xi}
\end{align}
where, $\mathbb{1}$ denotes the identity matrix, and the functions $\Pi_{i}^{\text{phy},\ H}(Q^2)$ (with $i=1,\dots,10$) are expressed in terms of the TFFs and the physical parameters of the baryons (see appendix~\ref{appA}).
It is important to emphasize that the physical side of the correlation function is identical for all spin-$3/2$ baryons. Furthermore, if we include the terms indicated by \textquotedblleft$\cdots$\textquotedblright\ in Eqs.~\eqref{phy omega}–\eqref{phy xi}, the physical side of the correlation function remains identical for all three baryons. For the numerical analysis, in Eqs.~\eqref{phy omega}–\eqref{phy xi}, we retain only the Lorentz structures that contribute to the TFFs. 
Contributions from higher states, the continuum, and other Lorentz structures are indicated by the dots.
	
	\subsection{QCD side}\label{QCD side}	
In this sector, the QCD side of the correlation function is evaluated in the large space-like momentum transfer region.
To this end, we insert the tensor currents from Eq.~\eqref{tensor-current} and the interpolating currents from Eq.~\eqref{J-ab} into Eq.~\eqref{corrf}. Considering the time-ordering operator in its definition, Wick’s theorem is then applied. 
Consequently, the correlation function is obtained on the QCD side in the form:
\begin{equation}
	\Pi_{\alpha\mu\nu\beta}^{\text{QCD},\ H}(p,p^\prime) = i^2 \varepsilon_{abc}\varepsilon_{a^\prime b^\prime c^\prime}
	\int d^4 x\ e^{-ip.x} \int d^4 y\ e^{ip'.y}\ \Pi^H_{\alpha\mu\nu\beta}(x,y),
	\label{corrf1}
\end{equation}	
where, $\Pi^H_{\alpha\mu\nu\beta}(x,y)$ for the selected baryon hyperons is formulated using gamma matrices and the corresponding light-quark propagators. 
Owing to the lengthy form of $\Pi^H_{\alpha\mu\nu\beta}(x,y)$, the complete expressions for the $\Omega^-$, $\Sigma^{*+}$, and $\Xi^{*-}$ baryons are collected in Appendix~\ref{appB}.
Using the light-quark propagator in Eq.~\eqref{propagator}, the correlation functions for the considered baryon hyperons are obtained from Eqs.~\eqref{iso omega}–\eqref{isosinglet sigma}.
\allowdisplaybreaks	
\begin{align}
	\Pi_{\alpha\mu\nu\beta}^{\text{QCD},\ H}(p,p')=\int d^4 x\ e^{-ip.x}\int d^4 y\ e^{ip'.y}&\Big[\Pi^{(\text{pert}),\ H}_{\alpha\mu\nu\beta}(p,p')+\Pi^{(3D),\ H}_{\alpha\mu\nu\beta}(p,p')+\Pi^{(4D),\ H}_{\alpha\mu\nu\beta}(p,p')
	\nonumber\\&+\Pi^{(5D),\ H}_{\alpha\mu\nu\beta}(p,p')+\Pi^{(6D),\ H}_{\alpha\mu\nu\beta}(p,p')+\cdots\Big],
	\label{corrf2}
\end{align}	
where, $D$ denotes the space-time dimension. The correlation function for the baryon hyperons receives a perturbative contribution, $\Pi^{(\text{pert}),\ H}_{\alpha\mu\nu\beta}(p,p')$, as well as non-perturbative contributions, $\Pi^{(3D),\ H}_{\alpha\mu\nu\beta}(p,p')$, $\Pi^{(4D),\ H}_{\alpha\mu\nu\beta}(p,p')$, $\Pi^{(5D),\ H}_{\alpha\mu\nu\beta}(p,p')$, and $\Pi^{(6D),\ H}_{\alpha\mu\nu\beta}(p,p')$, corresponding to operators of mass dimensions three through six, respectively.
Contributions from operators of higher mass dimensions are indicated by the dots. The terms with mass dimensions $0$, $3$, and $4$, and separately those with dimensions $5$ and $6$, were evaluated using two independent methods.
	
	\vspace{0.3cm}
	{\textbf{(i) The correlation function with mass dimensions $0$, $3$ and $4$:}} 
	\\
	\\
	The contributions to the correlation function with mass dimensions $0$, $3$, and $4$ are first computed in coordinate space and then transformed to momentum space for further evaluation using \cite{Azizi:2017ubq},
	\begin{equation}
		\dfrac{1}{(L^2)^{m_j}} = \int \frac{d^D p_j}{(2 \pi)^D} \exp{-ip_j.L} 
		\,i (-1)^{m_j + 1} 2^{D - 2 m_j} \pi^{D/2} 
		\dfrac{\Gamma[D/2 - m_j]}{\Gamma[m_j]} {\Big(-\frac{1}{p_j^2}\Big)}^{D/2 - m_j},
		\label{fourie}
	\end{equation}
	where $L$ stands for $x-y$, $x$, or $y$. The space-time coordinates are expressed as derivatives with respect to the momenta: $y_\mu = -i {\partial}/{\partial p'_\mu}$ and $x_\mu = i {\partial}/{\partial p_\mu}$.
	Performing the integrations over $x$ and $y$ in $D$-dimension generates two Dirac delta functions in momentum space, enabling the direct evaluation of two of the $D$-dimensional momentum integrals in Eq.~\eqref{fourie}. The remaining $D$-dimensional integral is then simplified via Feynman parametrization and subsequently computed using the formula: \cite{Dehghan:2023ytx, Azizi:2017ubq},
	\begin{equation}
		\int d^D\ell \frac{1}{(\ell^2 + \Delta)^n} = 
		\dfrac{i \pi^{D/2} (-1)^{n}\Gamma[n-D/2]}{\Gamma[n] (-\Delta)^{n-D/2}}.
		\label{integral}
	\end{equation}
	Within the dispersion relation framework, the invariant amplitudes corresponding to the different Lorentz structures on the QCD side can be written as double integrals over the spectral densities in momentum space,
	\allowdisplaybreaks
	\begin{align}
\Pi_{i}^{\text{QCD},\ H}(s_0, Q^2) = \int_{(m_{\psi_1}+m_{\psi_2}+m_{\psi_3})^2}^{s_0} ds \int_{(m_{\psi_1}+m_{\psi_2}+m_{\psi_3})^2}^{s_0} ds' \frac{\rho_i^{H}(s,s',Q^2)}{(s-p^2)(s'-p'^2)},
	\label{spectral density}
	\end{align}
	the parameter $s_0$ corresponds to the continuum threshold and takes distinct values for each baryon hyperon. Because the initial and final states are the same hyperon, the lower and upper bounds of the double integrals coincide.
	The spectral densities $\rho_i^H(s,s',Q^2)$ are obtained from the imaginary parts of the QCD side invariant functions,
	\allowdisplaybreaks
	\begin{align}
\rho_i^H(s,s',Q^2) =\dfrac{\text{Im}[\Pi_{i}^{\text{QCD},\ H}(s_0,Q^2)]}{\pi}.
	\end{align}
The imaginary parts corresponding to the different Lorentz structures can be computed using\cite{Azizi:2017ubq},
	\begin{equation}
		\Gamma[\frac{D}{2} - n] {\Big(\frac{-1}{\Delta}\Big)}^{D/2 - n} = 
		\frac{(-1)^{n-1}}{(n-2)!} (-\Delta)^{n-2} \ln [-\Delta].
		\label{imarinarypart}
	\end{equation}
	Subsequently, we work in $D=4$ space-time dimensions. Following the same procedure as on the physical side in \ref{physical side}, the gamma matrices are arranged in the prescribed order, and the constraints necessary to eliminate the spin-$1/2$ contamination are then imposed.
	
	\vspace{0.3cm}
	{\textbf{(ii) The correlation function with mass dimensions $5$ and  $6$:}}\\
	\\
	For these contributions, we first rotate to Euclidean space via a Wick rotation, and subsequently employ the Schwinger parametrization as:
	\begin{equation}
		\dfrac{1}{(L^2)^{m_j}}= \frac{(-1)^{m_j}}{\Gamma(m_j)} \int_0^\infty d\alpha\, \alpha^{m_j-1} \exp{\left[ -\alpha L^2\right] },
		\label{Schwinger}
	\end{equation}
	where $\alpha$ denotes an independent Schwinger parameter. We then carry out the Gaussian integration over the Euclidean coordinates $x_E$ and $y_E$, yielding expressions in terms of the Schwinger parameters.
	 After setting $D=4$, a Wick rotation back to Minkowski space is performed, with $x_\mu$ and $y_\mu$ represented as derivatives with respect to the momenta.
	  Following the procedure on the physical side, the Dirac matrices are arranged similarly, and terms corresponding to spin-$1/2$ contamination are removed.
	   Subsequently, a double Borel transformation is applied with respect to the squared momenta of the initial and final baryons, using the Borel parameters $M_1^2$ and $M_2^2$ \cite{Asmaee:2025elo}.

	 The Schwinger parameter integrals are subsequently carried out. Applying the double Borel transformation to part 
	 \textbf{(i)} and performing continuum subtraction for both \textbf{(i)} and \textbf{(ii)} \cite{Ozdem:2017jqh, Azizi:2018duk}, we arrive at the final Borel-transformed expression of the correlation function on the QCD side:
	 \allowdisplaybreaks
	 \begin{align}
\Pi_i^{\text{QCD},\ H}(s_0, M^2, Q^2) = \int_{(m_{\psi_1}+m_{\psi_2}+m_{\psi_3})^2}^{s_0} ds \int_{(m_{\psi_1}+m_{\psi_2}+m_{\psi_3})^2}^{s_0} ds' \rho_i^{H}(s,s',Q^2)\ e^{-\frac{s}{2M^2}}\ e^{-\frac{s^\prime}{2M^2}}+\Gamma_i^{H}(s_0,M^2, Q^2),
	 \end{align}
in the above equation, the first term accounts for the contribution from part $\textbf{(i)}$, while the second term, $\Gamma_i^H(s_0,M^2, Q^2)  $, arises from part $\textbf{(ii)}$.
	The full QCD side correlation function for the baryon hyperons $\Omega^-$, $\Sigma^{*+}$, and $\Xi^{*-}$ in the Borel scheme is then given by:
\allowdisplaybreaks
\begin{equation}
\begin{split}
	\Pi_{\alpha\mu\nu\beta}^{\text{QCD},\ \Omega^-}(s_0, M^2, Q^2) &=  g_{\alpha\beta}\gamma_{\mu}\gamma_\nu\slashed{p}^\prime \slashed{p}\Pi_1^{\text{QCD},\ \Omega^-}(s_0, M^2, Q^2)
	+g_{\alpha\mu}g_{\nu\beta}\slashed{p}^\prime\Pi_2^{\text{QCD},\ \Omega^-}(s_0, M^2, Q^2)
	\\&
	+ p_\alpha  \gamma_{\mu}\gamma_\nu p^\prime_\beta\slashed{p}^\prime \slashed{p}\Pi_3^{\text{QCD},\ \Omega^-}(s_0, M^2, Q^2)
	+p'_\mu g_{\alpha\beta}p_\nu\slashed{p}\Pi_4^{\text{QCD},\ \Omega^-}(s_0, M^2, Q^2)
	\\&
	+p_\alpha g_{\mu\beta}\gamma_\nu \slashed{p}^\prime\Pi_5^{\text{QCD},\ \Omega^-}(s_0, M^2, Q^2)
	+p_\alpha p^\prime_\mu g_{\nu\beta} \slashed{p}^\prime\Pi_6^{\text{QCD},\ \Omega^-}(s_0, M^2, Q^2)
	\\&
	+p_\alpha p_\mu p^\prime_\nu p^\prime_\beta \slashed{p}^\prime\Pi_{7}^{\text{QCD},\ \Omega^-}(s_0, M^2, Q^2)
	+p_{\mu}g_{\alpha\beta}p^\prime_\nu \mathbb{1}\Pi_{8}^{\text{QCD},\ \Omega^-}(s_0, M^2, Q^2)
	\\&
	+p_\alpha p_\mu \gamma_\nu p^\prime_\beta \slashed{p} \Pi_{9}^{\text{QCD},\ \Omega^-}(s_0, M^2, Q^2)
	+p_\alpha p^\prime_\mu \gamma_\nu p^\prime_\beta \slashed{p}\Pi_{10}^{\text{QCD},\ \Omega^-}(s_0, M^2, Q^2)
	+\cdots,
	\label{QCD omega}
\end{split}
\end{equation}
	
	\allowdisplaybreaks
	\begin{align}
		\Pi_{\alpha\mu\nu\beta}^{\text{QCD},\ \Sigma^{*+}}(s_0, M^2, Q^2) &= g_{\alpha\beta}\gamma_{\mu}\gamma_\nu\slashed{p}^\prime \slashed{p}\Pi_1^{\text{QCD},\ \Sigma^{*+}}(s_0, M^2, Q^2)
		+g_{\alpha\mu}g_{\nu\beta}\slashed{p}^\prime \slashed{p}\Pi_2^{\text{QCD},\ \Sigma^{*+}}(s_0, M^2, Q^2)
		\nonumber\\
		&
		+ p_\alpha \gamma_{\mu}\gamma_\nu p^\prime_\beta\slashed{p}^\prime \slashed{p}\Pi_3^{\text{QCD},\ \Sigma^{*+}}(s_0, M^2, Q^2)
    	+g_{\alpha\beta} \gamma_\mu p^\prime_\nu\slashed{p}^\prime \slashed{p}\Pi_4^{\text{QCD},\ \Sigma^{*+}}(s_0, M^2, Q^2)
		\nonumber\\
     	&
		+g_{\alpha\beta} p_\mu p^\prime_\nu \slashed{p}^\prime\slashed{p}\Pi_5^{\text{QCD},\ \Sigma^{*+}}(s_0, M^2, Q^2)
		+p_\alpha g_{\mu\beta}\gamma_\nu \slashed{p}^\prime\slashed{p}\Pi_6^{\text{QCD},\ \Sigma^{*+}}(s_0, M^2, Q^2)
		\nonumber\\
		&+p_\alpha p_\mu g_{\nu\beta} \slashed{p}^\prime\slashed{p}\Pi_7^{\text{QCD},\ \Sigma^{*+}}(s_0, M^2, Q^2)
		+p_\alpha p'_\mu p_\nu p'_\beta  \mathbb{1} \Pi_8^{\text{QCD},\ \Sigma^{*+}}(s_0, M^2, Q^2)
		\nonumber\\
		&
		+p_\alpha p_\mu\gamma_\nu p^\prime_\beta \slashed{p}\Pi_{9}^{\text{QCD},\ \Sigma^{*+}}(s_0, M^2, Q^2)
		+p_\alpha p'_\mu\gamma_\nu p^\prime_\beta \slashed{p}\Pi_{10}^{\text{QCD},\ \Sigma^{*+}}(s_0, M^2, Q^2)
		+\cdots,
		\label{QCD sigma}
	\end{align}
	
	\allowdisplaybreaks
\begin{equation}
		\begin{split}
		\Pi_{\alpha\mu\nu\beta}^{\text{QCD},\ \Xi^{*-}}(s_0, M^2, Q^2)& = 
		g_{\alpha\beta}\gamma_{\mu}\gamma_\nu\slashed{p}^\prime \slashed{p}\Pi_1^{\text{QCD},\ \Xi^{*-}}(s_0, M^2, Q^2)
		+g_{\alpha\mu}g_{\nu\beta}\slashed{p}^\prime \slashed{p}\Pi_2^{\text{QCD},\ \Xi^{*-}}(s_0, M^2, Q^2)
		\\
		&
		+ p_\alpha \gamma_{\mu}\gamma_\nu p^\prime_\beta\slashed{p}^\prime \slashed{p}\Pi_3^{\text{QCD},\ \Xi^{*-}}(s_0, M^2, Q^2)
		+g_{\alpha\beta} \gamma_\mu p^\prime_\nu\slashed{p}^\prime \slashed{p}\Pi_4^{\text{QCD},\ \Xi^{*-}}(s_0, M^2, Q^2)
		\\
		&
		+g_{\alpha\beta} p_\mu p^\prime_\nu \slashed{p}^\prime\slashed{p}\Pi_5^{\text{QCD},\ \Xi^{*-}}(s_0, M^2, Q^2)
		+p_\alpha g_{\mu\beta}\gamma_\nu \slashed{p}^\prime\slashed{p}\Pi_6^{\text{QCD},\ \Xi^{*-}}(s_0, M^2, Q^2)
		\\
		&
		+p_\alpha p'_\mu g_{\nu\beta} \slashed{p}^\prime\Pi_7^{\text{QCD},\ \Xi^{*-}}(s_0, M^2, Q^2)
		+p_\alpha p'_\mu p_\nu p'_\beta  \mathbb{1}\Pi_{8}^{\text{QCD},\ \Xi^{*-}}(s_0, M^2, Q^2)
		\\
		&
		+p_\alpha p_\mu\gamma_\nu p^\prime_\beta \slashed{p}\Pi_{9}^{\text{QCD},\ \Xi^{*-}}(s_0, M^2, Q^2)
		+p_\alpha p'_\mu\gamma_\nu p^\prime_\beta \slashed{p}\Pi_{10}^{\text{QCD},\ \Xi^{*-}}(s_0, M^2, Q^2)
		+\cdots,
		\label{QCD xi}
	\end{split}
\end{equation}
		The functions $\Pi^{\text{QCD},\ H}_i(s_0, M^2, Q^2)$ for $H=\Omega^-, \Sigma^{*+}, \Xi^{*-}$, are quite lengthy and are not shown explicitly.
	The TFFs of the $\Omega^-$, $\Sigma^{*+}$, and $\Xi^{*-}$ baryons are determined by equating the coefficients of the chosen Lorentz structures on the QCD side, given by Eqs.~\eqref{QCD omega}–\eqref{QCD xi}, with those on the physical side from Eqs.~\eqref{phy omega}–\eqref{phy xi}.
	
	\section{ NUMERICAL ANALYSES}\label{ANALYSES}	
		In this section, we perform a numerical analysis of the TFFs obtained from the QCDSR formalism, using the input parameters summarized in Table \ref{input}.
	\begin{table}[!htb]
		\centering
		\begin{minipage}{0.90\textwidth} 
			\centering
			\renewcommand{\arraystretch}{1.5} 
			\setlength{\tabcolsep}{12pt}      
			\begin{tabular}{|c|c|c|c|c|c|}
				\hline
				\text{Input parameters} & \text{Values} & \text{Input parameters} & \text{Values} \\
				\hline\hline
				$m_s$ & $0.0935\pm 0.0005\ \text{GeV}$ \cite{ParticleDataGroup:2024cfk}  & $\lambda_{\Omega^-}$ & $0.049\pm0.015\ \mathrm{GeV}^3$ \cite{Aliev:2016jnp}\\
				\hline
				$m_u$ & $0.00216\pm0.00007\ \text{GeV}$ \cite{ParticleDataGroup:2024cfk}  & $\lambda_{\Sigma^{*+}}$ &$0.036\pm0.010\ \mathrm{GeV}^3$\cite{Aliev:2016jnp}\\
				\hline
				$m_d$ & $0.00470\pm0.00004\ \text{GeV}$ \cite{ParticleDataGroup:2024cfk} & $\lambda_{\Xi^{*-}}$ & $0.045\pm0.013\ \mathrm{GeV}^3$ \cite{Aliev:2016jnp}\\
				\hline
				$m_{\Omega^-}$ & $1.67243\pm 0.00032\ \text{GeV}$  \cite{ParticleDataGroup:2024cfk} & $\alpha_s$ &$(0.118 \pm 0.005)$ \cite{DELPHI:1993ukk}  \\
				\hline
				$m_{\Sigma^{*+}}$ & $1.38283\pm0.00034\ \text{GeV}$  \cite{ParticleDataGroup:2024cfk} &$g_s$& $\sqrt{4\pi \alpha_s}$ \\
				\hline
				$m_{\Xi^{*-}}$ & $1.5350\pm0.0006\ \text{GeV}$ \cite{ParticleDataGroup:2024cfk} & $\langle \frac{\alpha_s}{\pi} G^2 \rangle $ & $(0.012\pm0.004)$ $~\mathrm{GeV}^4 $ \cite{Belyaev:1982cd}\\
				\hline
				$\left\langle \bar{q}q \right\rangle,\ $\text{with}\ ($q=u,d,s$) & $(-0.24\pm 0.01)^3$ $\mathrm{GeV}^3$ \cite{Belyaev:1982sa} & $\left\langle \bar{q} g_s \sigma G q\right\rangle$ & $(0.8 \pm 0.1)\left\langle \bar{q}q \right\rangle \ \mathrm{GeV}^5$ \cite{Belyaev:1982sa}\\
				\hline
			\end{tabular}
		\end{minipage}
		\caption{Input parameters and corresponding values used in the numerical evaluations.}
		\label{input}
	\end{table}
	
	In the QCDSR analysis, aside from the standard input parameters, two additional auxiliary quantities: the Borel mass $M^2$ and the continuum threshold $s_0$, play a crucial role in the stability and reliability of the QCDSR.
	The auxiliary parameters are introduced as part of the computational procedure. Ideally, the extracted physical quantities should be independent of these parameters; however, a residual dependence remains in practice. The working intervals are chosen to satisfy three main criteria: stability of the TFFs under variations of the Borel mass $M^2$ and continuum threshold $s_0$, dominance of the ground state pole contribution (PC) over the excited states and continuum, and reliable convergence of the operator product expansion (OPE). The following conditions are imposed to ensure these requirements are met \cite{Asmaee:2025elo,Aliev:2016jnp}:
	\allowdisplaybreaks
	\begin{equation}
		\begin{split}
			&	\text{PC}(Q^2) = \frac{\Pi_{i}^{\text{QCD},\ H}(s_0, M^2, Q^2)}{\Pi_{i}^{\text{QCD},\ H}(\infty, M^2, Q^2)} \geqslant 0.5,\quad
			\text{R} (M^2, Q^2) = \frac{\Pi_{i}^{\text{QCD,\ H,\ \text{Dim}6}}(s_0, M^2, Q^2)}{\Pi_{i}^{\text{QCD},\ H}(\infty, M^2, Q^2)} \leqslant 0.08,
		\end{split}
	\end{equation}
	where, $i$ labels the selected Lorentz structures. The upper bound of the Borel parameter $M^2$ is fixed by the PC requirement, which ensures that the ground-state pole contributes at least 50\% of the total QCD amplitude. The lower bound is determined from the convergence of the OPE, limiting the contribution of the highest-dimensional operator to no more than 8\% of the total. Using these criteria, the appropriate working ranges for $s_0$ and $M^2$ are obtained and summarized in Table \ref{s0M2}.
	\begin{table}[!htb]
	\centering
	\begin{minipage}{0.90\textwidth} 
		\centering
		\renewcommand{\arraystretch}{1.5} 
		\setlength{\tabcolsep}{12pt}      
		\begin{tabular}{|c|c|c|}
			\hline
			\text{Baryon hyperons} & $s_0\ \big[\mathrm{GeV}^2\big] $ & $M^2\ \big[\mathrm{GeV}^2\big] $ \\
			\hline\hline
	       	$\Omega^-$ & $3.30\leq s_0\leq 3.50$ & $7\leq M^2\leq 8$\\
			\hline
			$\Sigma^{*+}$ & $3.24\leq s_0\leq 3.44$ & $5\leq M^2\leq 6$\\
			\hline
			$\Xi^{*-}$ & $4.00\leq s_0\leq 4.20$ & $5.5\leq M^2\leq 6.5$ \\
			\hline
		\end{tabular}
	\end{minipage}
	\caption{Working ranges of the continuum threshold $s_0$ and Borel parameter $M^2$ for the baryon hyperons $\Omega^-$, $\Sigma^{*+}$, and $\Xi^{*-}$.}
	\label{s0M2}
\end{table}
	
\begin{figure}[!htb]
		\centering
		\includegraphics[width=0.41\textwidth]{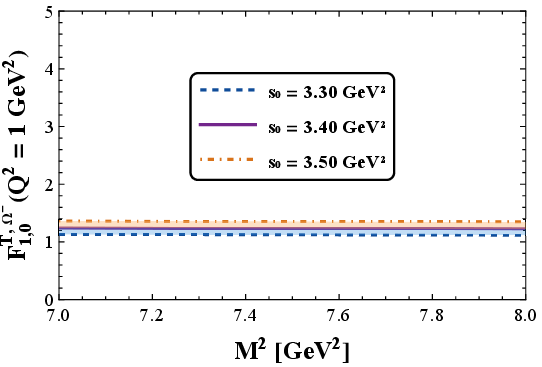}~~~~~~~~
		\includegraphics[width=0.41\textwidth]{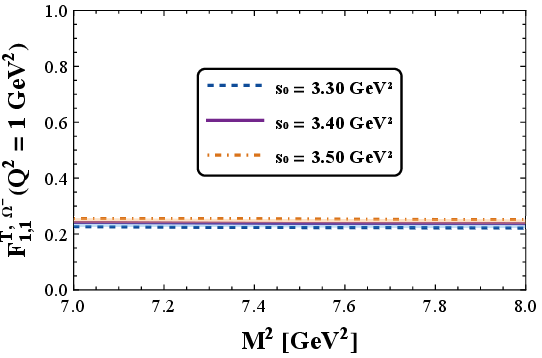}
		
		\vspace{0.2cm}
		\includegraphics[width=0.41\textwidth]{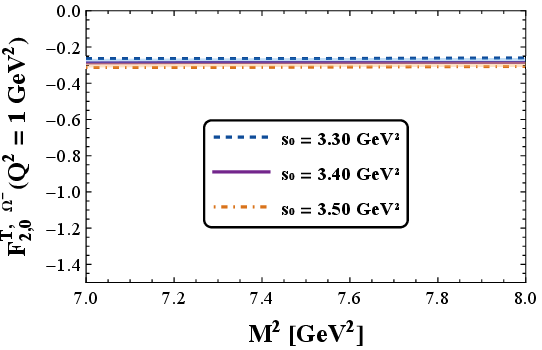}~~~~~~~~
		\includegraphics[width=0.41\textwidth]{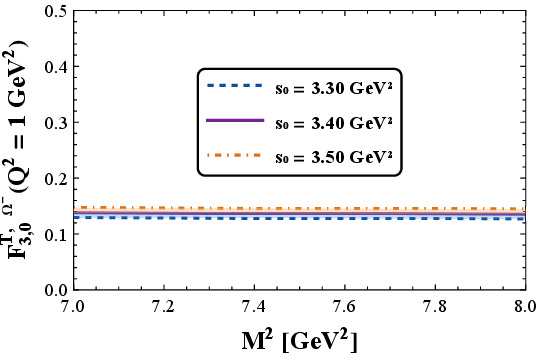}
		
		\vspace{0.2cm}
		\includegraphics[width=0.41\textwidth]{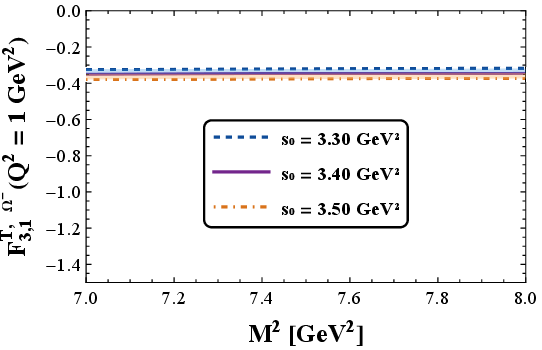}~~~~~~~~
		\includegraphics[width=0.41\textwidth]{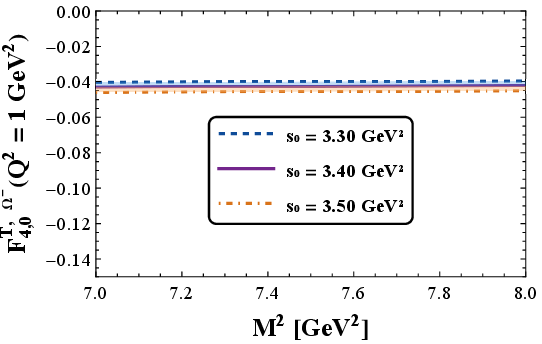}
		
		\vspace{0.2cm}
		\includegraphics[width=0.41\textwidth]{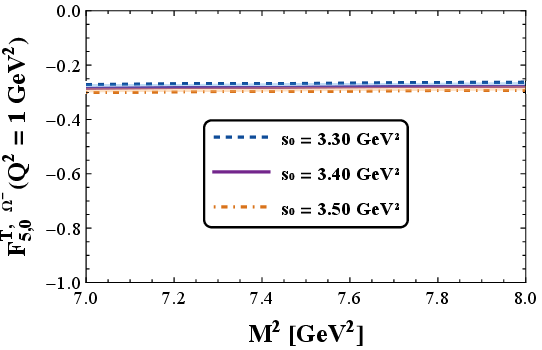}~~~~~~~~
		\includegraphics[width=0.41\textwidth]{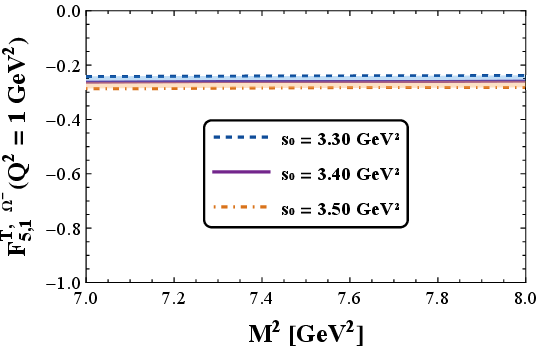}
		
		\vspace{0.2cm}
		\includegraphics[width=0.41\textwidth]{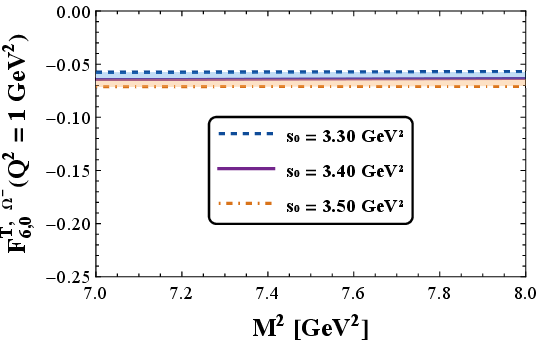}~~~~~~~~
		\includegraphics[width=0.41\textwidth]{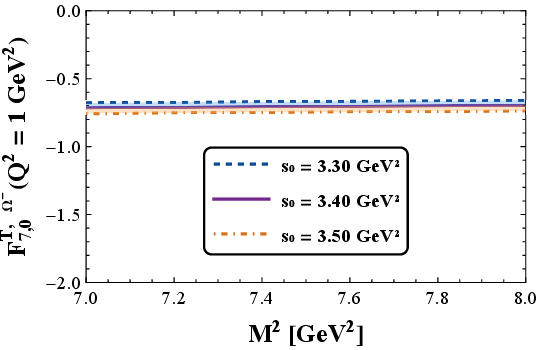}
		\caption{Dependence of the $\Omega^-$ TFFs on the Borel mass parameter $M^2$ at $Q^2 = 1.0~\text{GeV}^2$ for three different values of the continuum threshold $s_0$.}
		\label{Msqomega}
	\end{figure}

\begin{figure}[!htb]
		\centering
		\includegraphics[width=0.41\textwidth]{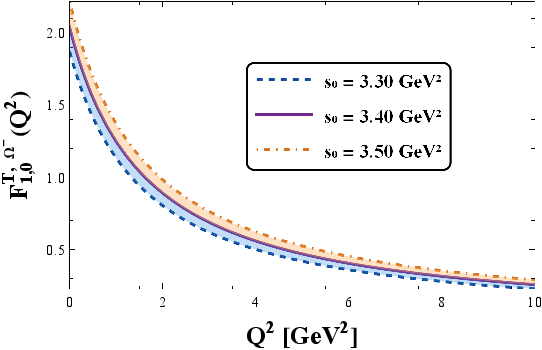}~~~~~~~~
		\includegraphics[width=0.41\textwidth]{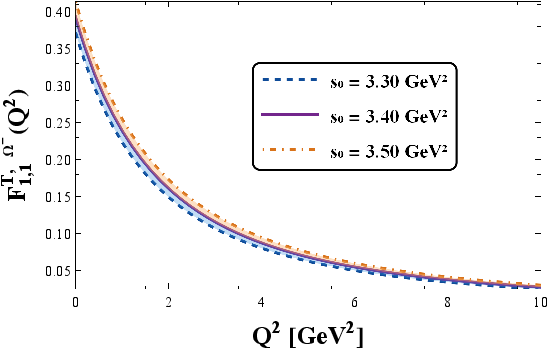}
		
		\vspace{0.2cm}
		\includegraphics[width=0.41\textwidth]{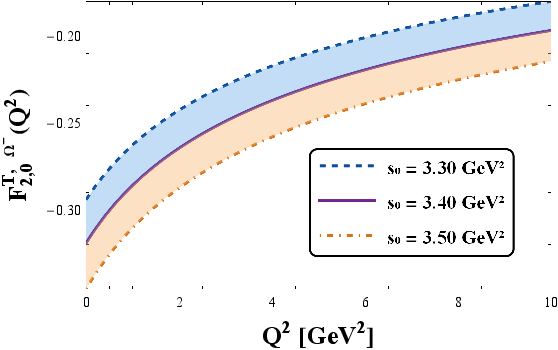}~~~~~~~~
		\includegraphics[width=0.41\textwidth]{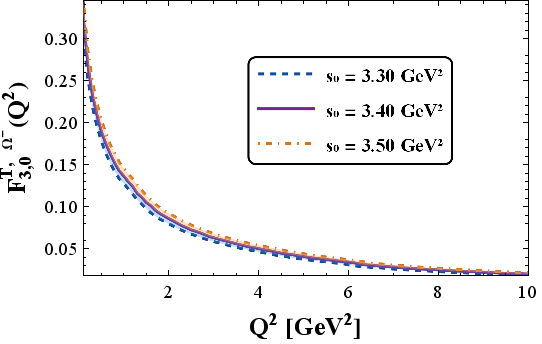}
		
		\vspace{0.2cm}
		\includegraphics[width=0.41\textwidth]{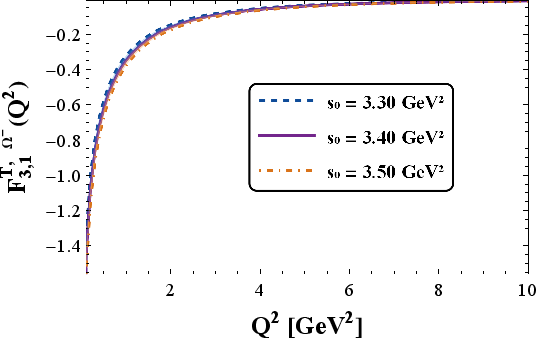}~~~~~~~~
		\includegraphics[width=0.41\textwidth]{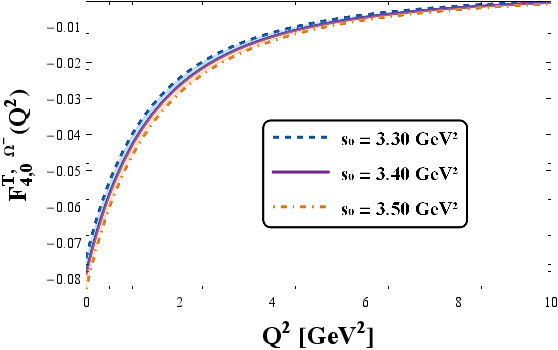}
		
		\vspace{0.2cm}
		\includegraphics[width=0.41\textwidth]{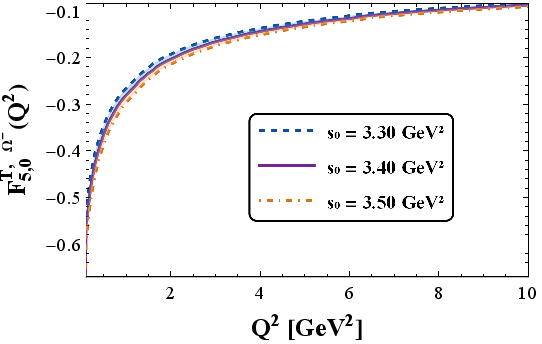}~~~~~~~~
		\includegraphics[width=0.41\textwidth]{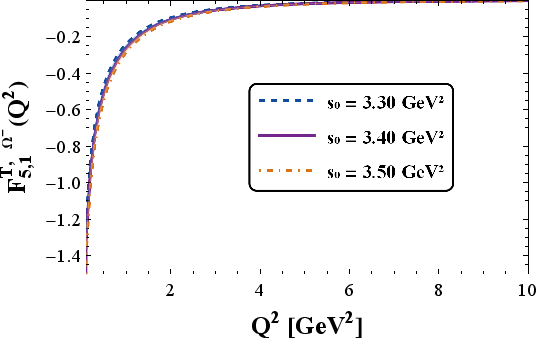}
		
		\vspace{0.2cm}
		\includegraphics[width=0.41\textwidth]{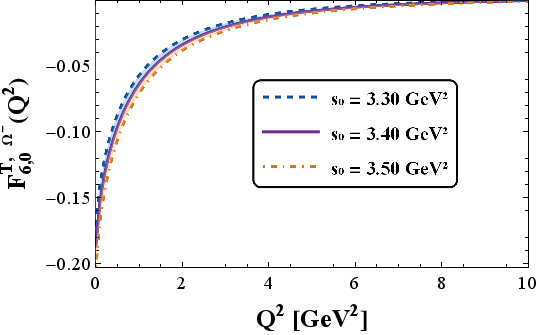}~~~~~~~~
		\includegraphics[width=0.41\textwidth]{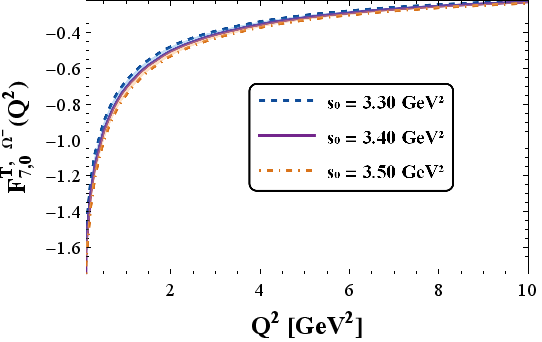}
		\caption{Variation of the $\Omega^-$ TFFs with $Q^2$ at a fixed Borel parameter $M^2 = 7.5~\text{GeV}^2$, shown for three choices of the continuum threshold $s_0$.}
		\label{Qsqomega}
	\end{figure}

\begin{figure}[!htb]
		\centering
		\includegraphics[width=0.41\textwidth]{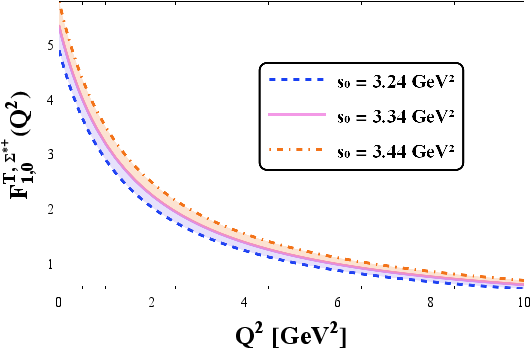}~~~~~~~~
		\includegraphics[width=0.41\textwidth]{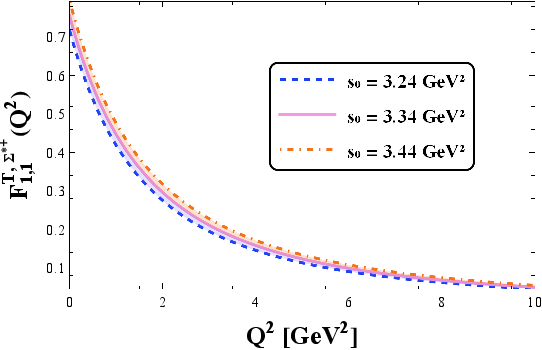}
		
		\vspace{0.2cm}
		\includegraphics[width=0.41\textwidth]{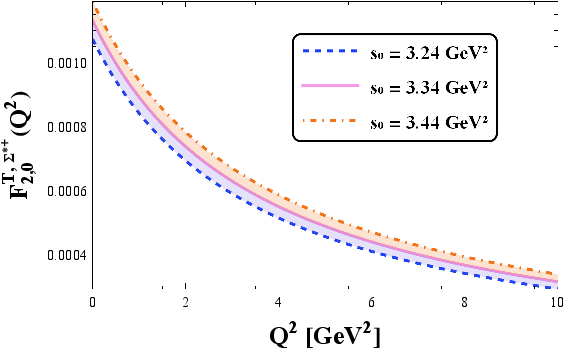}~~~~~~~~
		\includegraphics[width=0.41\textwidth]{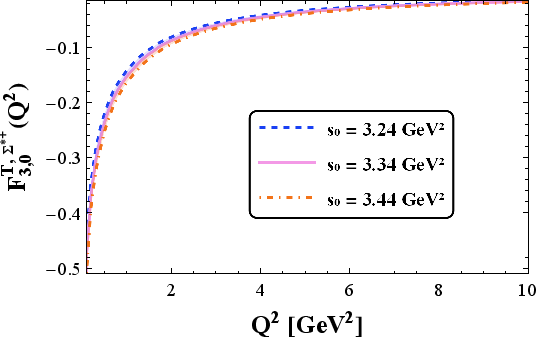}
		
		\vspace{0.2cm}
		\includegraphics[width=0.41\textwidth]{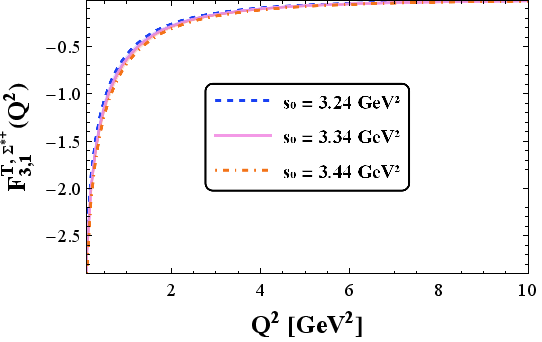}~~~~~~~~
		\includegraphics[width=0.41\textwidth]{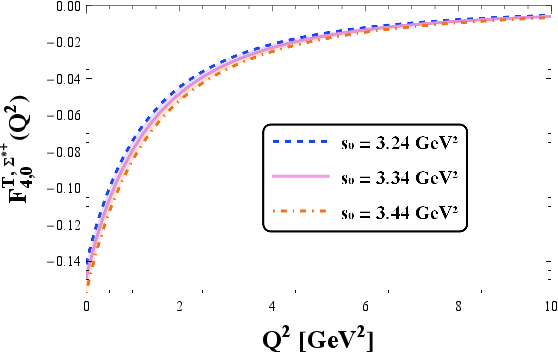}
		
		\vspace{0.2cm}
		\includegraphics[width=0.41\textwidth]{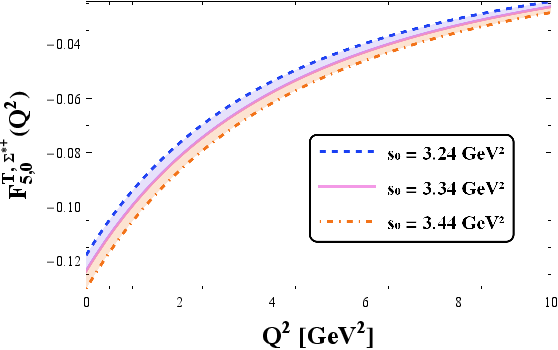}~~~~~~~~
		\includegraphics[width=0.41\textwidth]{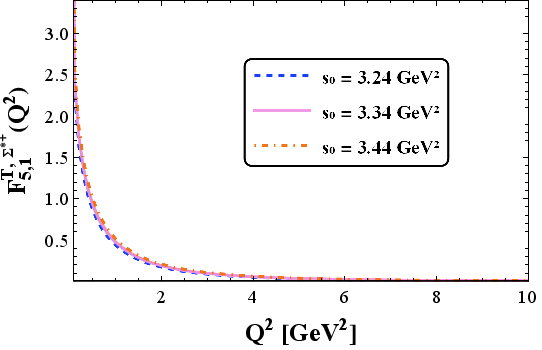}
		
		\vspace{0.2cm}
		\includegraphics[width=0.41\textwidth]{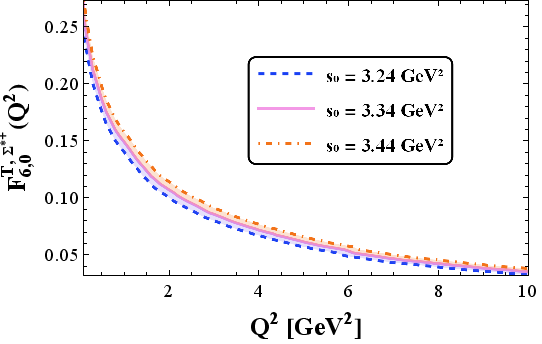}~~~~~~~~
		\includegraphics[width=0.41\textwidth]{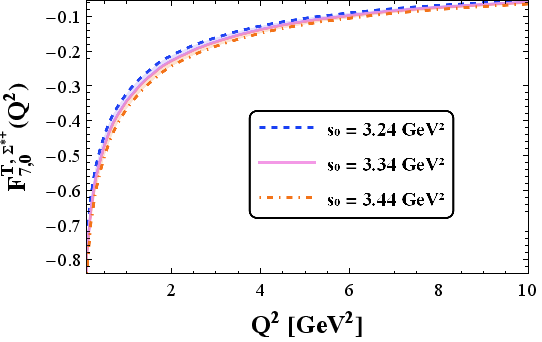}
		\caption{Variation of the $\Sigma^{*+}$ TFFs with $Q^2$ at $M^2 = 5.5~\text{GeV}^2$ for three choices of the continuum threshold $s_0$.}
		\label{Qsqisosingletsigma}
	\end{figure}

\begin{figure}[!htb]
		\centering
		\includegraphics[width=0.41\textwidth]{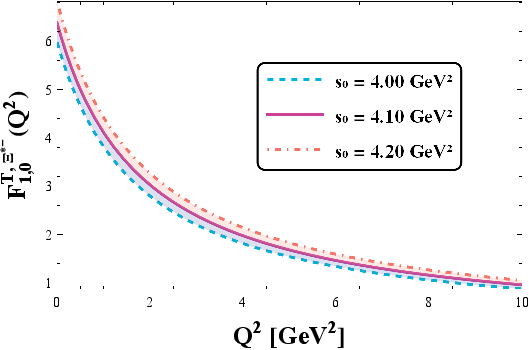}~~~~~~~~
		\includegraphics[width=0.41\textwidth]{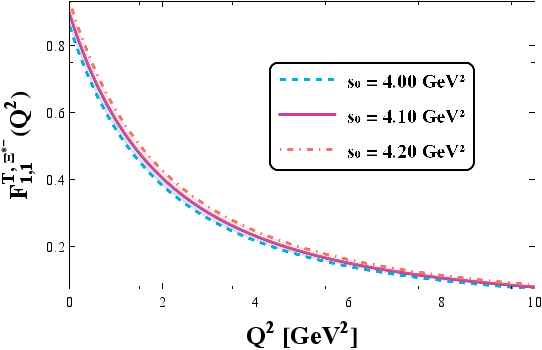}
		
		\vspace{0.2cm}
		\includegraphics[width=0.41\textwidth]{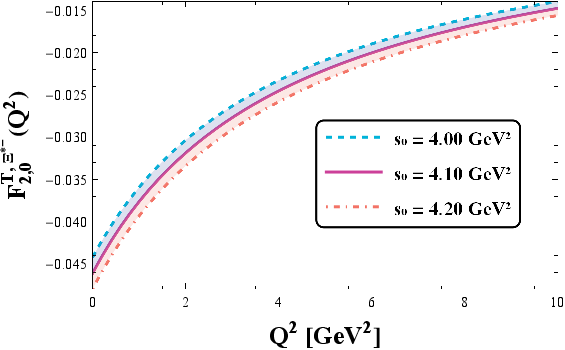}~~~~~~~~
		\includegraphics[width=0.41\textwidth]{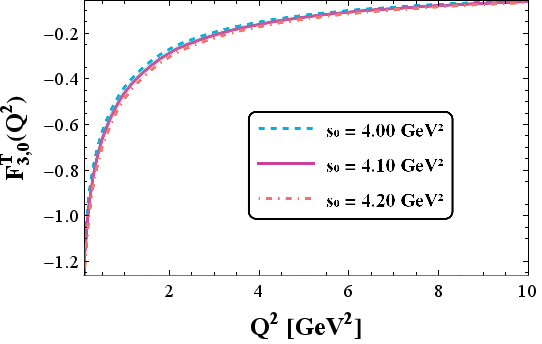}

		\vspace{0.2cm}
		\includegraphics[width=0.41\textwidth]{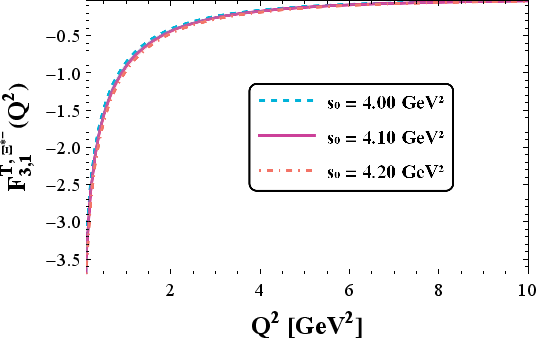}~~~~~~~~
		\includegraphics[width=0.41\textwidth]{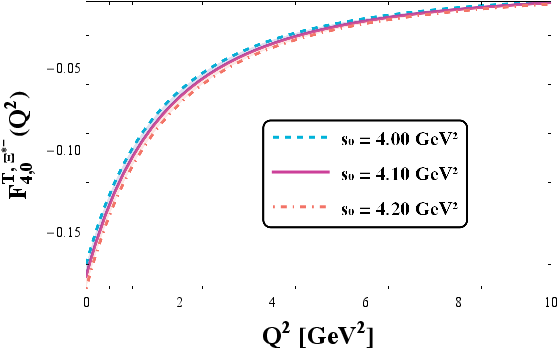}

		\vspace{0.2cm}
		\includegraphics[width=0.41\textwidth]{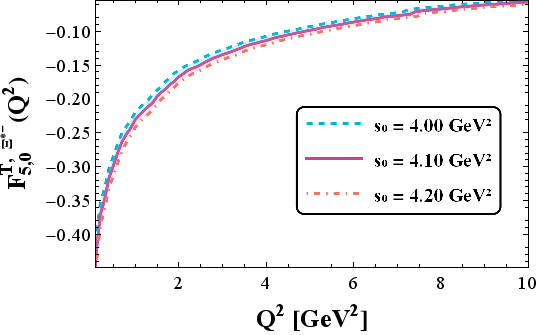}~~~~~~~~
		\includegraphics[width=0.41\textwidth]{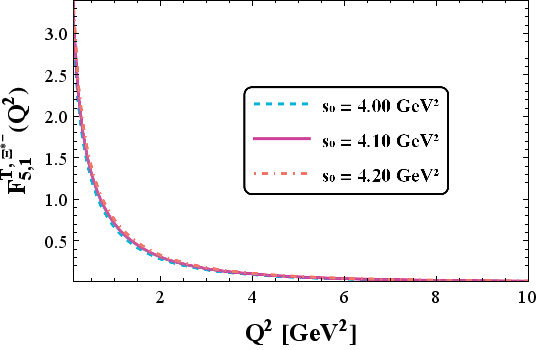}
		
		\vspace{0.2cm}
		\includegraphics[width=0.41\textwidth]{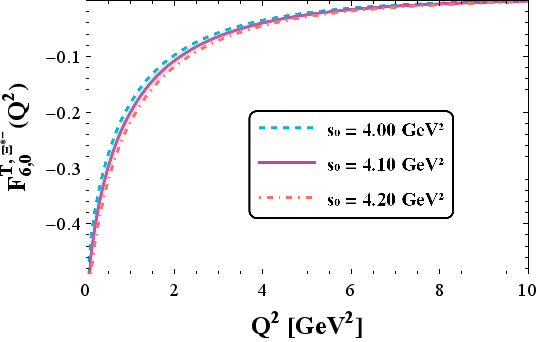}~~~~~~~~
		\includegraphics[width=0.41\textwidth]{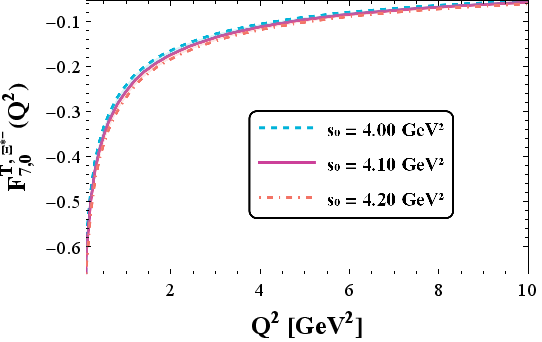}
		\caption{Variation of the $\Xi^{*-}$ TFFs at $M^2 = 6.0~\text{GeV}^2$ for three choices of $s_0$.}
		\label{Qsqisovectorsigma}
	\end{figure}

The dependence of the TFFs on the Borel parameter $M^2$ and the squared momentum transfer $Q^2$ is illustrated in Figs.~\ref{Msqomega}–\ref{Qsqisovectorsigma} for the baryon hyperons $\Omega^-$, $\Sigma^{*+}$, and $\Xi^{*-}$.
	The dependence of the TFFs on the Borel parameter $M^2$ is analyzed at $Q^2 = 1.0~\text{GeV}^2$ for three representative values of the continuum threshold $s_0$ for all considered baryon hyperons. For brevity, the corresponding results are presented only for the $\Omega^-$ baryon as a representative example in Fig.~\ref{Msqomega}.
	The results demonstrate that the TFFs exhibit only minor variations within the chosen Borel window, indicating a stable and reliable determination from the QCDSR approach.
The $Q^2$ dependence of the TFFs has been analyzed for all considered baryon hyperons. In particular, Fig.~\ref{Qsqomega} illustrates the $Q^2$ dependence of the TFFs for the $\Omega^-$ baryon at a fixed Borel mass parameter
$M^2 = 7.5\ \text{GeV}^2$ and three continuum thresholds, $s_0 = 3.30, 3.40$, and $3.50\ \text{GeV}^2$. Figs.~\ref{Qsqisosingletsigma} and \ref{Qsqisovectorsigma} show the $Q^2$ dependence of the TFFs for $\Sigma^{*+}$ at $M^2 = 5.5\ \text{GeV}^2$ (with $s_0 = 3.24, 3.34,$ and $3.44\ \text{GeV}^2$) and for $\Xi^{*-}$ at $M^2 = 6.0\ \text{GeV}^2$ (with $s_0 = 4.00, 4.10,$ and $4.20\ \text{GeV}^2$), respectively.

	The QCDSR results for the $\Omega^-$, $\Sigma^{*+}$, and $\Xi^{*-}$ hyperons are accurately captured by the generalized $\mathbf{p}$-pole parametrization \cite{Dehghan:2023ytx, Asmaee:2025elo}
	\begin{equation}
		{\cal F}(Q^2)=\frac{{\cal F}(0)}{\left(1+\frac{Q^2}{m_\mathbf{p}^2} \right)^\mathbf{p} },
		\label{dipole fit}
	\end{equation}
	in this expression, $m_\mathbf{p}$ serves as an effective mass scale in GeV, while $\mathbf{p}$ and ${\cal F}(0)$ are dimensionless fit parameters. The $\mathbf{p}$-pole parametrization effectively describes the $Q^2$ dependence of the TFFs for the baryon hyperons, exhibiting a gradual decrease and approaching zero around $Q^2 \sim 10\ \text{GeV}^2$, as shown in Figs.~\ref{Qsqomega}, \ref{Qsqisosingletsigma}, and \ref{Qsqisovectorsigma}.
		\begin{table}[!htb]
		\centering
		\begin{minipage}{0.90\textwidth} 
			\centering
			\renewcommand{\arraystretch}{1.42} 
			\setlength{\tabcolsep}{12pt}      
			\begin{tabular}{|c|c|c|c|}
				\hline
				TFF & ${\cal F}(0)$ & $m_\mathbf{p}$ & $\mathbf{p}$ \\
				\hline\hline
				$F^{T,\ \Omega^-}_{1,0}(Q^2)$ & $2.04 \pm 0.15$ & $1.32 \pm 0.00$ & $1.09 \pm 0.02$ \\
				\hline
				$F^{T,\ \Omega^-}_{1,1}(Q^2)$ & $0.39 \pm 0.02$ & $1.87 \pm 0.02$ & $1.97 \pm 0.01$ \\
				\hline
				$F^{T,\ \Omega^-}_{2,0}(Q^2)$ & $-0.32 \pm 0.02$ & $1.40 \pm 0.02$ & $0.27 \pm 0.00$ \\
				\hline
				$F^{T,\ \Omega^-}_{3,0}(Q^2)$ & $0.43 \pm 0.19$ & $0.72 \pm 0.00$ & $0.94 \pm 0.01$ \\
				\hline
				$F^{T,\ \Omega^-}_{3,1}(Q^2)$ & $-1.92 \pm 0.10$ & $0.67 \pm 0.00$ & $1.51 \pm 0.01$ \\
				\hline
				$F^{T,\ \Omega^-}_{4,0}(Q^2)$ & $-0.06 \pm 0.05$ & $1.75 \pm 0.02$ & $2.17 \pm 0.01$ \\
				\hline
				$F^{T,\ \Omega^-}_{5,0}(Q^2)$ & $-0.67 \pm 0.03$ & $0.50 \pm 0.01$ & $0.55 \pm 0.01$ \\
				\hline
				$F^{T,\ \Omega^-}_{5,1}(Q^2)$ & $-1.80 \pm 0.10$ & $0.68 \pm 0.00$ & $1.71 \pm 0.01$ \\
				\hline
				$F^{T,\ \Omega^-}_{6,0}(Q^2)$ & $-0.16 \pm 0.01$ & $1.54 \pm 0.02$ & $2.63 \pm 0.11$ \\
				\hline
				$F^{T,\ \Omega^-}_{7,0}(Q^2)$ & $-2.11 \pm 0.08$ & $0.37 \pm 0.01$ & $0.52 \pm 0.01$ \\
				\hline
			\end{tabular}
		\end{minipage}
		\caption{Numerical values of the $\mathbf{p}$-pole fit parameters for the TFFs of the $\Omega^-$ baryon (Fig.~\ref{Qsqomega}), obtained at the central values of the continuum threshold $s_0$ and Borel mass $M^2$.}
		\label{fitparameters-omega}
	\end{table}
	
	\begin{table}[!htb]
		\centering
		\begin{minipage}{0.90\textwidth}
			\centering
			\renewcommand{\arraystretch}{1.42} 
			\setlength{\tabcolsep}{12pt}      
			\begin{tabular}{|c|c|c|c|}
				\hline
				TFF & ${\cal F}(0)$ & $m_\mathbf{p}$ & $\mathbf{p}$ \\
				\hline\hline
				$F^{T,\ \Sigma^{*+}}_{1,0}(Q^2)$ & $5.33 \pm 0.39$ & $1.31 \pm 0.00$ & $1.12 \pm 0.02$ \\
				\hline
				$F^{T,\ \Sigma^{*+}}_{1,1}(Q^2)$ & $0.76 \pm 0.03$ & $1.78 \pm 0.02$ & $1.91 \pm 0.01$ \\
				\hline
				$F^{T,\ \Sigma^{*+}}_{2,0}(Q^2)$ & $0.00 \pm 0.00$ & $1.75 \pm 0.34$ & $0.94 \pm 0.01$ \\
				\hline
				$F^{T,\ \Sigma^{*+}}_{3,0}(Q^2)$ & $-0.51 \pm 0.03$ & $0.58 \pm 0.00$ & $1.01 \pm 0.00$ \\
				\hline
				$F^{T,\ \Sigma^{*+}}_{3,1}(Q^2)$ & $-3.63 \pm 0.18$ & $0.67 \pm 0.00$ & $1.53 \pm 0.01$ \\
				\hline
				$F^{T,\ \Sigma^{*+}}_{4,0}(Q^2)$ & $-0.15 \pm 0.01$ & $1.72 \pm 0.02$ & $2.18 \pm 0.01$ \\
				\hline
				$F^{T,\ \Sigma^{*+}}_{5,0}(Q^2)$ & $-0.10 \pm 0.08$ & $2.99 \pm 0.08$ & $2.07 \pm 0.05$ \\
				\hline
				$F^{T,\ \Sigma^{*+}}_{5,1}(Q^2)$ & $3.41 \pm 0.17$ & $0.67 \pm 0.22$ & $1.73 \pm 0.01$ \\
				\hline
				$F^{T,\ \Sigma^{*+}}_{6,0}(Q^2)$ & $0.27 \pm 0.01$ & $0.95 \pm 0.00$ & $0.80 \pm 0.01$ \\
				\hline
				$F^{T,\ \Sigma^{*+}}_{7,0}(Q^2)$ & $-0.90 \pm 0.04$ & $0.68 \pm 0.02$ & $0.84 \pm 0.00$ \\
				\hline
			\end{tabular}
			\caption{Numerical values of the $\mathbf{p}$-pole fit parameters for the TFFs of the $\Sigma^{*+}$ baryon (Fig.~\ref{Qsqisosingletsigma}), evaluated at the central values of the continuum threshold $s_0$ and Borel mass $M^2$.}
			\label{fitparameters-isoSsigma}
		\end{minipage}
	\end{table}
	
	\begin{table}[!htb]
		\centering
		\begin{minipage}{0.90\textwidth}
			\centering
			\renewcommand{\arraystretch}{1.42} 
			\setlength{\tabcolsep}{12pt}      
			\begin{tabular}{|c|c|c|c|}
				\hline
				TFF & ${\cal F}(0)$ & $m_\mathbf{p}$ & $\mathbf{p}$ \\
				\hline\hline
				$F^{T,\ \Xi^{*-}}_{1,0}(Q^2)$ & $6.39 \pm 0.36$ & $1.36 \pm 0.01$ & $1.02 \pm 0.01$ \\
				\hline
				$F^{T,\ \Xi^{*-}}_{1,1}(Q^2)$ & $0.90 \pm 0.03$ & $1.95 \pm 0.02$ & $1.89 \pm 0.01$ \\
				\hline
				$F^{T,\ \Xi^{*-}}_{2,0}(Q^2)$ & $-0.05 \pm 0.02$ & $1.99 \pm 0.02$ & $0.90 \pm 0.00$ \\
				\hline
				$F^{T,\ \Xi^{*-}}_{3,0}(Q^2)$ & $-1.40 \pm 0.06$ & $0.67 \pm 0.00$ & $0.95 \pm 0.01$ \\
				\hline
				$F^{T,\ \Xi^{*-}}_{3,1}(Q^2)$ & $-3.55 \pm 0.09$ & $0.69 \pm 0.00$ & $1.44 \pm 0.01$ \\
				\hline
				$F^{T,\ \Xi^{*-}}_{4,0}(Q^2)$ & $-0.18 \pm 0.01$ & $1.88 \pm 0.02$ & $2.14 \pm 0.00$ \\
				\hline
				$F^{T,\ \Xi^{*-}}_{5,0}(Q^2)$ & $-0.35 \pm 0.30$ & $0.83 \pm 0.00$ & $0.74 \pm 0.00$ \\
				\hline
				$F^{T,\ \Xi^{*-}}_{5,1}(Q^2)$ & $4.24\pm 0.18$ & $0.69 \pm 0.00$ & $1.62 \pm 0.01$ \\
				\hline
				$F^{T,\ \Xi^{*-}}_{6,0}(Q^2)$ & $-0.49 \pm 0.03$ & $1.68 \pm 0.01$ & $2.92 \pm 0.07$ \\
				\hline
				$F^{T,\ \Xi^{*-}}_{7,0}(Q^2)$ & $-0.78 \pm 0.03$ & $0.48 \pm 0.00$ & $0.67 \pm 0.00$ \\
				\hline
			\end{tabular}
			\caption{Numerical values of the $\mathbf{p}$-pole fit parameters for the  TFFs of the $\Xi^{*-}$ baryon (Fig.~\ref{Qsqisovectorsigma}), evaluated at the central values of the continuum threshold $s_0$ and Borel mass $M^2$.}
			\label{fitparameters-isoSxi}
		\end{minipage}
	\end{table}

The corresponding $\mathbf{p}$-pole parameters are listed in Tables~\ref{fitparameters-omega}–\ref{fitparameters-isoSxi}. These fits are obtained at the central values of $s_0$ and $M^2$, with the uncertainties reflecting both the dependence on the auxiliary parameters and the intrinsic systematic effects of the QCDSR framework.
	 A detailed analysis of the resulting TFFs provides valuable insight into the distribution of the tensor charge among the constituent quarks.
	The quark tensor charge is extracted in the forward limit ($q = 0$) \cite{Fu:2024kfx, Fu:2025upc}.
	Applying Eq.~\eqref{martix-element} under this condition yields \cite{Asmaee:2025elo}:
	\allowdisplaybreaks
	\begin{align}
		\left\langle H(p,s)\left|  \bar{\psi}(0)i\sigma_{\mu\nu}\psi(0)\right| H(p,s)\right\rangle=\bar{u}^\alpha(p,s)
		\Big[i\sigma_{\mu\nu} g_{\alpha\beta}F^{T,\ H}_{1,0}(0) 
		+ig_{\alpha [\mu}\sigma _{\nu]\beta}F_{2,0}^{T,\ H}(0)
		\Big]u^\beta(p,s).
	\end{align}
	By employing the Rarita–Schwinger constraints,
	\allowdisplaybreaks
	\begin{align}
		&\gamma_{\beta}u^\beta(p,s)=0,\quad p_\beta u^\beta(p,s)=0,
	\end{align}
	and taking into account the antisymmetric nature of $\sigma_{\mu\nu}$, we arrive at:
	\allowdisplaybreaks
	\begin{align}
		\left\langle H(p,s)\left|  \bar{\psi}(0)i\sigma_{\mu\nu}\psi(0)\right| H(p,s)\right\rangle=\bar{u}^\alpha(p,s) i\sigma_{\mu\nu} {u}_\alpha(p,s)
		\Big(F^{T,\ H}_{1,0}(0) -F_{2,0}^{T,\ H}(0)\Big).
	\end{align}
	The quark tensor charge, $\delta\psi^H$, is defined by the combination of TFFs,
	$F_{1,0}^{T,\ H}(0) - F_{2,0}^{T,\ H}(0)$, which represents the only non-vanishing
	contribution at zero momentum transfer.
	This result aligns with the findings of Ref.~\cite{Fu:2024kfx}.

	 Employing the 
	$\mathbf{p}$-pole fit parameters given in Tables~\ref{fitparameters-omega}–\ref{fitparameters-isoSxi}, the quark tensor charges for the hyperon baryons considered here are then calculated as follows:
	\begin{align}
		\begin{array}{l@{\hspace{2cm}}l@{\hspace{2cm}}l} 
			\delta\psi^{\Omega^-}=2.36\pm0.15,
			\\
			\\
			\delta\psi^{\Sigma^{*+}}=5.33\pm0.39
			\\
			\\
			\delta\psi^{\Xi^{*-}}=6.44\pm0.36.
		\end{array}
		\label{tensor charge}
	\end{align}	
For the baryon hyperons $\Omega^-$, $\Sigma^{*+}$, and $\Xi^{*-}$, the TFFs provide a detailed probe of their internal spin structure and the distribution of quark transversity. Similar to the $\Delta^+$ baryon case \cite{Asmaee:2025elo}, the quark tensor charge extracted in the forward limit reflects the intrinsic spin correlations of the constituent quarks. 
Beyond the tensor charge, the full set of TFFs encodes rich information about the internal structure of spin-$3/2$ baryons. They provide quantitative measures of the spin distribution among the quarks, characterize electromagnetic properties induced by tensor interactions, and may even offer indirect insights into the spatial and geometric structure of these baryons.
These results are not only relevant for understanding the intrinsic properties of hyperons within the Standard Model, but also serve as essential inputs for constructing observables sensitive to the beyond Standard Model physics, such as searches for quark electric dipole moments or tensor interactions in low-energy experiments. The TFFs thus act as fundamental building blocks for both theoretical studies and potential experimental probes of spin dependent hadronic structure \cite{Fu:2024kfx, Pospelov:2005pr, Wang:2018kto}.
	\section{Summary and conclusion}\label{conclusion}
	Hadrons interact through a variety of currents, and each type of interaction is characterized by a distinct set of FFs. These FFs encode critical information about the internal structure, spin distributions, and dynamic properties of hadrons. While experimental studies have provided some insights into hadronic interactions, discrepancies between theoretical predictions and experimental measurements persist. These differences highlight the complexity of hadron structure and indicate that our understanding remains incomplete.
	For baryon hyperons in particular, theoretical and experimental information is still scarce. This limited knowledge restricts our ability to fully comprehend their internal dynamics, interactions, and intrinsic properties. Previous studies have primarily focused on their electromagnetic and gravitational FFs, offering only a partial understanding of their structure. To achieve a more complete picture, it is essential to investigate their interactions with additional currents, such as the tensor current, which provides complementary insights into quark spin correlations and the distribution of tensor charges within these baryons.
	The present study contributes to this effort by analyzing the TFFs of the $\Omega^-$, $\Sigma^{*+}$, and $\Xi^{*-}$ baryons within the framework of QCDSR. Our results provide new theoretical predictions that can guide future experimental investigations and help refine models of decuplet baryon structure. Ultimately, a comprehensive understanding of these FFs is crucial for advancing our knowledge of hadronic physics and for testing the underlying principles of QCD in the non-perturbative regime.
	TFFs of baryon hyperons are important because they help us understand the spin structure of these baryons and show detailed properties, such as the distribution of tensor charge. These FFs, obtained from the matrix elements of the tensor current between hadron states, provide a useful way to study the shape and internal dynamics of baryons and play a key role in understanding the non-perturbative features of QCD.
	With recent progress in studying the interactions of these currents, especially tensor and gravitational currents with hadrons, more precise experimental data are expected  soon. These new data will not only help clarify the properties of baryon hyperons but also significantly improve our understanding of the structure and dynamics of nucleons.
	This data serve as important input for experiments at leading research centers around the world, such as the LHC, JLab, Mainz, LAPP, and other advanced laboratories. These centers can provide valuable information on the interactions of various currents with nucleons and baryon hyperons.
	In this work, we have investigated hyperon–hyperon transitions induced by the tensor current within the framework of QCDSR. The QCD side of the correlation function was computed for the baryon hyperons, and the corresponding Lorentz structures were systematically matched to those on the physical side. This procedure allowed us to derive a complete set of ten independent TFFs for the baron hyperons $\Omega^-$, $\Sigma^{*+}$, and $\Xi^{*-}$.
	The resulting sum rules were analyzed numerically over the momentum transfer range $Q^2 \in [0,10\ \mathrm{GeV}^2]$. The QCDSR approach, being a fully relativistic and non-perturbative method, enables a consistent treatment of spin-$3/2$ baryons and incorporates the essential features of QCD dynamics, including quark and gluon condensates. Our analysis demonstrates that the $Q^2$ dependence of the extracted TFFs is well described by a generalized \textbf{p}-pole parametrization, from which we also determine reliable values of the FFs at $Q^2=0$.
	The results presented here provide new theoretical insight into the tensor structure of baryon hyperons, for which both experimental data and theoretical predictions remain limited. Our findings can serve as valuable benchmarks for future phenomenological studies and lattice QCD calculations. Moreover, ongoing and planned experimental programs, such as those at JLab have identified the study of spin-$3/2$ baryons and their GPDs as important objectives. In this context, the TFFs obtained in this work constitute essential building blocks for a deeper understanding of the internal structure, spin correlations, and non-perturbative dynamics of hyperons in QCD.
\newpage
\appendix
\renewcommand{\thesection}{\Alph{section}}
\renewcommand{\thesubsection}{\thesection. \arabic{subsection}}	
\section{The functions of physical side} \label{appA}
In this appendix, we provide the functions $\Pi_i^{\text{phy},\ H}(Q^2)$ for the baryon hyperons $\Omega^-$, $\Sigma^{*+}$, and $\Xi^{*-}$, which depend on the TFFs as well as the physical parameters appearing in the hadronic representation of the correlation function.

\vspace{0.5cm}
\underline{\large \textbf{$\Omega^-$ baryon:}}
\allowdisplaybreaks
\begin{align}
	\begin{array}{l@{\hspace{2cm}}l@{\hspace{2cm}}l} 
		\Pi_1^{\text{phy},\ \Omega^-}(Q^2)=F_{1,0}^{T,\ \Omega^-}(Q^2), 
		&
		\Pi_{6}^{\text{phy},\ \Omega^-}(Q^2)=\dfrac{1}{m_{\Omega^-}}F^{T,\ \Omega^-}_{7,0}(Q^2),
		\\
		\\
		\Pi_2^{\text{phy},\ \Omega^-}(Q^2)=-m_{\Omega^-}F_{2,0}^{T,\ \Omega^-}(Q^2),
		&
		\Pi_{7}^{\text{phy},\ \Omega^-}(Q^2)=-\dfrac{2}{m_{\Omega^-}^3}\left(F^{T,\ \Omega^-}_{3,1}(Q^2)-2 F^{T,\ \Omega^-}_{4,0}(Q^2)+F^{T,\ \Omega^-}_{5,1}(Q^2)\right),
		\\
		\\
		\Pi_3^{\text{phy},\ \Omega^-}(Q^2)=-\dfrac{1}{m_{\Omega^-}^2}F^{T,\ \Omega^-}_{1,1}(Q^2),
		&
		\Pi_8^{\text{phy},\ \Omega^-}(Q^2)=2F_{3,0}^{T,\ \Omega^-}(Q^2)+2F_{5,0}^{T,\ \Omega^-}(Q^2),
		\\
		\\
		\Pi_{4}^{\text{phy},\ \Omega^-}(Q^2)=-\dfrac{2}{m_{\Omega^-}^2}F^{T,\ \Omega^-}_{5,0}(Q^2),
		&
		\Pi_{9}^{\text{phy},\ \Omega^-}(Q^2)=\dfrac{1}{m_{\Omega^-}^2}\left(F^{T,\ \Omega^-}_{3,1}(Q^2)+2F^{T,\ \Omega^-}_{4,0}(Q^2) \right) ,
		\\
		\\
		\Pi_5^{\text{phy},\ \Omega^-}(Q^2)=F^{T,\ \Omega^-}_{6,0}(Q^2), 
		&
		\Pi_{10}^{\text{phy},\ \Omega^-}(Q^2)=\dfrac{1}{m_{\Omega^-}^2}\left(-F^{T,\ \Omega^-}_{3,1}(Q^2)
		+2F^{T,\ \Omega^-}_{4,0}(Q^2) \right).
	\end{array}
	\label{function omega}
\end{align}

\underline{\large \textbf{$\Sigma^{*+}$ and $\Xi^{*-}$ baryons:}	}
\begin{align}
	\begin{array}{l@{\hspace{2cm}}l@{\hspace{2cm}}l} 
		\Pi_1^{\text{phy},\ H'}(Q^2)=F_{1,0}^{T,\ H'}(Q^2), 
		&
		\Pi_6^{\text{phy},\ H'}(Q^2)=-\dfrac{1}{m_{H'}}F^{T,\ H'}_{6,0}(Q^2), 
		\\
		\\
		\Pi_2^{\text{phy},\ H'}(Q^2)=-F_{2,0}^{T,\ H'}(Q^2),
		&
		\Pi_{7}^{\text{phy},\ \Sigma^{*+}}(Q^2)=-\dfrac{1}{m_{\Sigma^{*+}}^2}F^{T,\ \Sigma^{*+}}_{7,0}(Q^2),\quad \Pi_{7}^{\text{phy},\ \Xi^{*-}}(Q^2)=\dfrac{1}{m_{\Xi^{*-}}}F^{T,\ \Xi^{*-}}_{7,0}(Q^2),
		\\
		\\
		\Pi_3^{\text{phy},\ H'}(Q^2)=-\dfrac{1}{m_{H'}^2}F^{T,\ H'}_{1,1}(Q^2),
			&
		\Pi_{8}^{\text{phy},\ H'}(Q^2)=\dfrac{2}{m_{H'}^2}\left(F^{T,\ H'}_{3,1}(Q^2)-2 F^{T,\ H'}_{4,0}(Q^2)+F^{T,\ H'}_{5,1}(Q^2)\right),
		\\
		\\
		\Pi_4^{\text{phy},\ H'}(Q^2)=-\dfrac{1}{m_{H'}}F^{T,\ H'}_{3,0}(Q^2),
		&
		\Pi_{9}^{\text{phy},\ H'}(Q^2)=\dfrac{1}{m_{H'}^2}\left(F^{T,\ H'}_{3,1}(Q^2)+2F^{T,\ H'}_{4,0}(Q^2) \right) , 
		\\
		\\
		\Pi_{5}^{\text{phy},\ H'}(Q^2)=\dfrac{2}{m_{H'}^2}F^{T,\ H'}_{5,0}(Q^2),
		&
		\Pi_{10}^{\text{phy},\ H'}(Q^2)=\dfrac{1}{m_{H'}^2}\left(-F^{T,\ H'}_{3,1}(Q^2)+2F^{T,\ H'}_{4,0}(Q^2) \right),
	\end{array}
	\label{function sigma}
\end{align}
where, $H'$ serves as a label for the $\Sigma^{*+}$ and $\Xi^{*-}$ baryons.

	\section{QCD side results} \label{appB}
	In this appendix, we provide the explicit expressions for $\Pi^H_{\alpha\mu\nu\beta}(x,y)$, formulated in terms of Dirac gamma matrices and the light-quark propagators for the baryon hyperons under consideration.

\vspace{0.3cm}
\underline{\large \textbf{$\Omega^-$ baryon:}	}
\allowdisplaybreaks
\begin{align}
	\Pi^{\Omega^-}_{\alpha\mu\nu\beta}(x, y)=&
	-\bigg[ S_s^{cb^\prime}(y-x)\gb S_s^{\prime\ ea^\prime}(0-x)\s S_s^{\prime\ ae}(y-0)\ga S_s^{bc^\prime}(y-x)\bigg]
	\nonumber\\
	&+\bigg[ S_s^{ca^\prime}(y-x)\gb S_s^{\prime\ eb^\prime}(0-x)\s S_s^{\prime\ ae}(y-0)\ga S_s^{bc^\prime}(y-x)\bigg]
	\nonumber\\
	&+\bigg[S_s^{c c^\prime}(y-x)\bigg] \text{Tr}\bigg[\ga S_s^{ae}(y-0) \s S_s^{e a^\prime}(0-x) \gb S_s^{\prime\ b b^\prime}(y-x)\bigg]
	\nonumber\\
	&-\bigg[S_s^{c a^\prime}(y-x)\gb S_s^{\prime\ b b^\prime}(y-x)\ga S_s^{ae}(y-0) \s S_s^{e c^\prime}(0-x)\bigg]
	\nonumber\\
	&-\bigg[S_s^{cc^\prime}(y-x)\bigg]\ \text{Tr}\bigg[\ga S_s^{ba^\prime}(y-x)\gb S_s^{\prime\ eb^\prime}(0-x)\s S_s^{\prime\ ae}(y-0)\bigg]
	\nonumber\\
	&+\bigg[S_s^{cb^\prime}(y-x)\gb S_s^{\prime\ ba^\prime}(y-x)\ga S_s^{ ae}(y-0)\s S_s^{ec^\prime}(0-x)\bigg]
	\nonumber\\
	&+\bigg[S_s^{cb^\prime}(y-x)\gb S_s^{\prime\ ea^\prime}(0-x)\s S_s^{\prime\ be}(y-0)\ga S_s^{ac^\prime}(y-x)\bigg]
	\nonumber\\
	&-\bigg[S_s^{ca^\prime}(y-x) \gb S_s^{\prime\ eb^\prime}(0-x)\s S_s^{\prime\ be}(y-0)\ga S_s^{ac^\prime}(y-x)\bigg]
	\nonumber\\
	&-\bigg[S_s^{ce}(y-0)\s S_s^{e a^\prime}(0-x)\gb S_s^{\prime\ b b^\prime}(y-x)\ga S_s^{a c^\prime}(y-x)\bigg]
	\nonumber\\
	&+\bigg[S_s^{ce}(y-0)\s S_s^{eb^\prime}(0-x)\gb S_s^{\prime\ ba^\prime}(y-x)\ga S_s^{ac^\prime}(y-x)\bigg]
	\nonumber\\
	&-\bigg[S_s^{cc^\prime}(y-x)\bigg]\ \text{Tr}\bigg[\ga S_s^{be}(y-0)\s S_s^{ ea^\prime}(0-x)\gb S_s^{\prime\ ab^\prime}(y-x)\bigg]
	\nonumber\\
	&+\bigg[ S_s^{ca^\prime}(y-x)\gb S_s^{\prime\ ab^\prime}(y-x)\ga S_s^{be}(y-0) \s S_s^{ec^\prime}(0-x)\bigg]
	\nonumber\\
	&+\bigg[S_s^{ce}(y-0) \s S_s^{ea^\prime}(0-x) \gb S_s^{\prime\ ab^\prime}(y-x)\ga S_s^{bc^\prime}(y-x)\bigg]
	\nonumber\\
	&-\bigg[S_s^{ce}(y-0)\s S_s^{ec^\prime}(0-x)\bigg]\ \text{Tr}\bigg[\ga S_s^{ba^\prime}(y-x)\gb S_s^{\prime\ ab^\prime}(y-x)\bigg]
	\nonumber\\
	&+\bigg[S_s^{cc^\prime}(y-x)\bigg]\ \text{Tr}\bigg[\ga S_s^{be}(y-0)\s S_s^{ eb^\prime}(0-x)\gb S_s^{\prime\ aa^\prime}(y-x)\bigg]
	\nonumber\\
	&-\bigg[S_s^{cb^\prime}(y-x)\gb S_s^{\prime\ aa^\prime}(y-x)\ga S_s^{be}(y-0)\s S_s^{ec^\prime}(0-x)\bigg]
	\nonumber\\
	&-\bigg[S_s^{ce}(y-0) \s S_s^{eb^\prime}(0-x) \gb S_s^{\prime\ aa^\prime}(y-x)\ga S_s^{bc^\prime}(y-x)\bigg]
	\nonumber\\
	&+\bigg[S_s^{ce}(y-0)\s S_s^{ec^\prime}(0-x)\bigg]\ \text{Tr}\bigg[\ga S_s^{aa^\prime}(y-x) \gb S_s^{\prime\ b b^\prime}(y-x)\bigg].
	\label{iso omega}
\end{align}

\underline{\large \textbf{$\Sigma^{*+}$ baryon:}	}
\allowdisplaybreaks
\begin{align}
	\Pi^{\Sigma^{*+}}_{\alpha\mu\nu\beta}(x,y)=\dfrac{1}{3}&
	\bigg(
	4\ \bigg[S_u^{c c^\prime}(y-x)\bigg] \text{Tr}\bigg[\ga S_u^{ae}(y-0) \s S_u^{e a^\prime}(0-x) \gb S_s^{\prime\ b b^\prime}(y-x)\bigg]
	\nonumber	\\
	&-4\bigg[S_u^{c a^\prime}(y-x)\gb S_s^{\prime\ b b^\prime}(y-x)\ga S_u^{ae}(y-0) \s S_u^{e c^\prime}(0-x)\bigg]
	\nonumber	\\
	&-4\bigg[S_u^{ce}(y-0)\s S_u^{e a^\prime}(0-x)\gb S_s^{\prime\ b b^\prime}(y-x)\ga S_u^{a c^\prime}(y-x)\bigg]
	\nonumber	\\
	&+4\bigg[S_u^{ce}(y-0)\s S_u^{ec^\prime}(0-x)\bigg]\ \text{Tr}\bigg[\ga S_u^{aa^\prime}(y-x) \gb S_s^{\prime\ b b^\prime}(y-x)\bigg]
	\nonumber	\\
	&-2\bigg[ S_u^{cb^\prime}(y-x)\gb S_u^{\prime\ ea^\prime}(0-x)\s S_u^{\prime\ ae}(y-0)\ga S_s^{bc^\prime}(y-x)\bigg]
	\nonumber	\\
	&+2\bigg[ S_u^{ca^\prime}(y-x)\gb S_u^{\prime\ eb^\prime}(0-x)\s S_u^{\prime\ ae}(y-0)\ga S_s^{bc^\prime}(y-x)\bigg]
	\nonumber	\\
	&+2\bigg[S_u^{ce}(y-0) \s S_u^{ea^\prime}(0-x) \gb S_u^{\prime\ ab^\prime}(y-x)\ga S_s^{bc^\prime}(y-x)\bigg]
	\nonumber	\\
	&-2\bigg[S_u^{ce}(y-0) \s S_u^{eb^\prime}(0-x) \gb S_u^{\prime\ aa^\prime}(y-x)\ga S_s^{bc^\prime}(y-x)\bigg]
	\nonumber	\\
	&-4\bigg[S_u^{ca^\prime}(y-x) \gb S_s^{\prime\ eb^\prime}(0-x)\s S_s^{\prime\ be}(y-0)\ga S_u^{ac^\prime}(y-x)\bigg]
	\nonumber	\\
	&+4\bigg[S_u^{cc^\prime}(y-x)\bigg]\ \text{Tr}\bigg[\ga S_s^{be}(y-0) \s S_s^{eb^\prime}(0-x) \gb S_u^{\prime\ aa^\prime}(y-x)\bigg]
	\nonumber	\\
	&+2\bigg[ S_u^{ca^\prime}(y-x)\gb S_u^{\prime\ ab^\prime}(y-x)\ga S_s^{be}(y-0) \s S_s^{ec^\prime}(0-x)\bigg]
	\nonumber	\\
	&-2\bigg[S_u^{cb^\prime}(y-x)\gb S_u^{\prime\ aa^\prime}(y-x)\ga S_s^{be}(y-0)\s S_s^{ec^\prime}(0-x)\bigg]
	\nonumber	\\
	&-2\bigg[S_s^{cb^\prime}(y-x)\gb S_u^{\prime\ ea^\prime}(0-x)\s S_u^{\prime\ ae}(y-0)\ga S_u^{bc^\prime}(y-x)\bigg]
	\nonumber	\\
	&+2\bigg[S_s^{cb^\prime}(y-x)\gb S_u^{\prime\ ba^\prime}(y-x)\ga S_u^{ ae}(y-0)\s S_u^{ec^\prime}(0-x)\bigg]
	\nonumber	\\
	&+2\bigg[S_s^{cb^\prime}(y-x)\gb S_u^{\prime\ ea^\prime}(0-x)\s S_u^{\prime\ be}(y-0)\ga S_u^{ac^\prime}(y-x)\bigg]
	\nonumber\\
	&-2\bigg[S_s^{cb^\prime}(y-x)\gb S_u^{\prime\ aa^\prime}(y-x)\ga S_u^{ be}(y-0)\s S_u^{ec^\prime}(0-x)\bigg]
	\nonumber	\\
	&+\bigg[S_s^{cc^\prime}(y-x)\bigg]\ \text{Tr}\bigg[\ga S_u^{bb^\prime}(y-x)\gb S_u^{\prime\ ea^\prime}(0-x)\s S_u^{\prime\ ae}(y-0)\bigg]
	\nonumber	\\
	&-\bigg[S_s^{cc^\prime}(y-x)\bigg]\ \text{Tr}\bigg[\ga S_u^{ba^\prime}(y-x)\gb S_u^{\prime\ eb^\prime}(0-x)\s S_u^{\prime\ ae}(y-0)\bigg]
	\nonumber	\\
	&-\bigg[S_s^{cc^\prime}(y-x)\bigg]\ \text{Tr}\bigg[\ga S_u^{be}(y-0)\s S_u^{ ea^\prime}(0-x)\gb S_u^{\prime\ ab^\prime}(y-x)\bigg]
	\nonumber	\\
	&+\bigg[S_s^{cc^\prime}(y-x)\bigg]\ \text{Tr}\bigg[\ga S_u^{be}(y-0)\s S_u^{ eb^\prime}(0-x)\gb S_u^{\prime\ aa^\prime}(y-x)\bigg]
	\nonumber	\\
	&+2\bigg[S_s^{ce}(y-0)\s S_s^{eb^\prime}(0-x)\gb S_u^{\prime\ ba^\prime}(y-x)\ga S_u^{ac^\prime}(y-x)\bigg]
	\nonumber	\\
	&-2\bigg[S_s^{ce}(y-0)\s S_s^{eb^\prime}(0-x)\gb S_u^{\prime\ aa^\prime}(y-x)\ga S_u^{bc^\prime}(y-x)\bigg]
	\nonumber	\\
	&-\bigg[S_s^{ce}(y-0)\s S_s^{ec^\prime}(0-x)\bigg]\ \text{Tr}\bigg[\ga S_u^{ba^\prime}(y-x)\gb S_u^{\prime\ ab^\prime}(y-x)\bigg]
	\nonumber	\\
	&+\bigg[S_s^{ce}(y-0) \s S_s^{ec^\prime}(0-x)\bigg]
	\text{Tr}\bigg[\ga S_u^{bb^\prime}(y-x) \gb S_u^{\prime\ aa^\prime}(y-x)\bigg]\bigg).
	\label{isosinglet sigma}
\end{align}

The expression for $\Pi^{\Xi^{*-}}_{\alpha\mu\nu\beta}$ is obtained from Eq.~\eqref{isosinglet sigma} by the replacements $u \rightarrow s$ and $s \rightarrow d$, and therefore we refrain from presenting it explicitly to avoid repetition.

In Eqs.~\eqref{iso omega}–\eqref{isosinglet sigma}, $S_\psi^{ab}(x)$ with $\psi=u, s$ denotes the propagator of a light-quark, and $S^{\prime\ ab}_\psi = C \big[ S^{ab}_\psi \big] ^TC$ represents the corresponding charge-conjugated propagator. The light-quark propagator $S_\psi^{ab}(x)$ in coordinate space, is expressed as \cite{Agaev:2020zad}:
\allowdisplaybreaks
	\begin{align}
	S_\psi^{ab}(x) &= i \delta^{ab} \frac{\slashed{x}}{2\pi^2 x^4} 
	- \delta^{ab} \frac{m_\psi}{4 \pi^2 x^2} 
	-\delta^{ab} \frac{\langle \bar{\psi} \psi \rangle}{12}\bigg(1-\frac{m_\psi \slashed{x}}{4}\bigg)
	- \delta^{ab} \frac{x^2}{192}\left\langle  \bar{\psi} g_s \sigma G \psi \right\rangle\bigg(1-i\frac{m_\psi \slashed{x}}{6}\bigg)
	\nonumber\\
	& \quad 
	- i \frac{g_s G^{ab}_{\alpha \beta}}{32 \pi^2 x^2} \left[ \slashed{x} \sigma^{\alpha \beta} + \sigma^{\alpha \beta} \slashed{x} \right]
	- i \delta^{ab} \frac{x^2 \slashed{x} g_s^2 \langle \bar{\psi} \psi \rangle^2}{7776} 
	+ \cdots,
	\label{propagator}
\end{align}
where, $m_\psi$ represents the mass of the light-quark, and  $g_s$ is the strong coupling constant. The quantities
$\langle \bar{\psi} \psi \rangle$, $\langle G^2 \rangle$ and $\langle \bar{\psi} g_s \sigma G \psi\rangle$ correspond to the quark, gluon, and mixed quark–gluon condensates, respectively. The gluon field-strength tensor
 $G^{ab}_{\alpha\beta}$ is,
\allowdisplaybreaks
\begin{align}
G_{\alpha\beta}^{ab} = G_{\alpha\beta}^{A} T_{A}^{ab}, \qquad 
T_{A} = \frac{1}{2} \lambda_{A}, \qquad
G^2 = G^{A}_{\alpha\beta} G^{A}_{\alpha\beta},
\end{align}
In the above expressions, $a, b = 1, 2, 3$ and $A = 1,\dots, 8$ represent the color indices of quark and gluon fields, respectively.
The SU($3$) generators in the fundamental representation are given by $T_A=\lambda_A/2$, where $\lambda_A$ are the Gell-Mann matrices. The gluon condensate can be expressed explicitly with its color and Lorentz indices as \cite{Asmaee:2025elo},
\allowdisplaybreaks
\begin{align}
\langle0|G^{ab}_{\alpha'\beta'}(0) G^{cd}_{\alpha\beta}(0)|0\rangle=\frac{\langle G^2 \rangle}{192} 
\left( g_{\alpha^\prime\alpha} g_{\beta^\prime\beta} - g_{\alpha^\prime\beta} g_{\beta^\prime\alpha} \right)
\left( \delta^{ad} \delta^{bc} - \frac{1}{3} \delta^{ab} \delta^{cd} \right).
\label{condensation gluon}
\end{align}

\begin{acknowledgments} 
	Z. Asmaee and K. Azizi are thankful to the Iran National Science Foundation (INSF) for the financial support
	provided for this research under the grant number 4036353.

\end{acknowledgments}


\begin{thebibliography}{999}
\bibitem{Hohler:1974eq}
G.~Hohler and E.~Pietarinen,
Electromagnetic Radii of Nucleon and Pion,
\href{https://doi.org//10.1016/0370-2693(75)90220-8}{Phys. Lett. B \textbf{53}, 471-475 (1975)}.

\bibitem{Maris:2000sk}
P.~Maris and P.~C.~Tandy,
The pi, K+, and K0 electromagnetic form-factors,
\href{https://doi.org//doi:10.1103/PhysRevC.62.055204}{Phys. Rev. C \textbf{62}, 055204 (2000)},
\href{https://arxiv.org/abs/nucl-th/0005015}{\color{teal}{[arXiv:nucl-th/0005015 [nucl-th]]}}.

\bibitem{Broniowski:2008hx}
W.~Broniowski and E.~Ruiz Arriola,
Gravitational and higher-order form factors of the pion in chiral quark models,
\href{https://doi.org//10.1103/PhysRevD.78.094011}{Phys. Rev. D \textbf{78}, 094011 (2008)},
\href{https://arxiv.org/abs/0809.1744}{\color{teal}{[arXiv:0809.1744 [hep-ph]]}}.

\bibitem{Er:2022cxx}
N.~Er and K.~Azizi,
Spectroscopic parameters and electromagnetic form factor of kaon in vacuum and a dense medium,
\href{https://doi.org//doi:10.1140/epjc/s10052-022-10333-w}{Eur. Phys. J. C \textbf{82}, 397 (2022)},
\href{https://arxiv.org/abs/2202.01504}{\color{teal}{[arXiv:2202.01504 [hep-ph]]}}.

\bibitem{Esmer:2025xss}
G.~D.~Esmer, K.~Azizi, H.~Sundu and S.~T{\"u}rkmen,
Decays of the light hybrid meson 1-+,
\href{https://doi.org//10.1103/PhysRevD.111.034041}{Phys. Rev. D \textbf{111}, 034041 (2025)},
\href{https://arxiv.org/abs/2501.11331}{\color{teal}{[arXiv:2501.11331 [hep-ph]]}}.

\bibitem{Federbush:1958zz}
P.~Federbush, M.~L.~Goldberger and S.~B.~Treiman,
Electromagnetic Structure of the Nucleon,
\href{https://doi.org//10.1103/PhysRev.112.642}{Phys. Rev. \textbf{112}, 642-665 (1958)}.

\bibitem{Bincer:1959tz}
A.~M.~Bincer,
Electromagnetic structure of the nucleon,
\href{https://doi.org//10.1103/PhysRev.118.855}{Phys. Rev. \textbf{118}, 855-863 (1960)}.

\bibitem{Ma:2002ir}
B.~Q.~Ma, D.~Qing and I.~Schmidt,
Electromagnetic form-factors of nucleons in a light cone diquark model,
\href{https://doi.org//10.1103/PhysRevC.65.035205}{Phys. Rev. C \textbf{65}, 035205 (2002)},
\href{https://arxiv.org/abs/hep-ph/0202015}{\color{teal}{[arXiv:hep-ph/0202015 [hep-ph]]}}.

\bibitem{Aliev:2008cs}
T.~M.~Aliev, K.~Azizi, A.~Ozpineci and M.~Savci,
Nucleon Electromagnetic Form Factors in QCD,
\href{https://doi.org//10.1103/PhysRevD.77.114014}{Phys. Rev. D \textbf{77}, 114014 (2008)},
\href{https://arxiv.org/abs/0802.3008 }{\color{teal}{[arXiv:0802.3008 [hep-ph]]}}.

\bibitem{Aliev:2011ku}
T.~M.~Aliev, K.~Azizi and M.~Savci,
Nucleon tensor form factors induced by isovector and isoscalar currents in QCD,
\href{https://doi.org//10.1103/PhysRevD.84.076005}{Phys. Rev. D \textbf{84}, 076005 (2011)},
\href{https://arxiv.org/abs/1108.2019 }{\color{teal}{[arXiv:1108.2019 [hep-ph]]}}.

\bibitem{Azizi:2015fqa}
K.~Azizi and H.~Sundu,
Radiative transition of negative to positive parity nucleon,
\href{https://doi.org//10.1103/PhysRevD.91.093012}{Phys. Rev. D \textbf{91}, 093012 (2015)},
\href{https://arxiv.org/abs/1501.07691}{\color{teal}{[arXiv:1501.07691 [hep-ph]]}}.


\bibitem{Chen:2022odn}
C.~Chen and C.~D.~Roberts,
Nucleon axial form factor at large momentum transfers,
\href{https://doi.org//10.1140/epja/s10050-022-00848-x}{Eur. Phys. J. A \textbf{58}, 206 (2022)},
\href{https://arxiv.org/abs/2206.12518}{\color{teal}{[arXiv:2206.12518 [hep-ph]]}}.

\bibitem{Dehghan:2025ncw}
Z.~Dehghan, F.~Almaksusi and K.~Azizi,
Mechanical properties of proton using flavor-decomposed gravitational form factors,
\href{https://doi.org//10.1007/JHEP06(2025)025}{JHEP \textbf{06}, 025 (2025)},
\href{https://arxiv.org/abs/2502.16689}{\color{teal}{[arXiv:2502.16689 [hep-ph]]}}.


\bibitem{Najjar:2024deh}
Z.~R.~Najjar, K.~Azizi and H.~R.~Moshfegh,
Properties of the ground and excited states of triply heavy spin-1/2 baryons,
\href{https://doi.org//10.1140/epjc/s10052-024-12960-x}{Eur. Phys. J. C \textbf{84}, 612 (2024)},
\href{https://arxiv.org/abs/2402.14348}{\color{teal}{[arXiv:2402.14348 [hep-ph]]}}.

\bibitem{ShekariTousi:2024mso}
M.~Shekari Tousi and K.~Azizi,
Properties of doubly heavy spin-12 baryons: The ground and excited states,
\href{https://doi.org//10.1103/PhysRevD.109.054005}{Phys. Rev. D \textbf{109}, 054005 (2024)},
\href{https://arxiv.org/abs/2401.07151}{\color{teal}{[arXiv:2401.07151 [hep-ph]]}}.


\bibitem{Dong:2013rk}
Y.~Dong and C.~Liang,
Generalized parton distribution functions of a deuteron in a phenomenological Lagrangian approach,
\href{https://doi.org//10.1088/0954-3899/40/2/025001}{J. Phys. G \textbf{40}, 025001 (2013)}.

\bibitem{Sun:2017gtz}
B.~D.~Sun and Y.~B.~Dong,
$\rho$ meson unpolarized generalized parton distributions with a light-front constituent quark model,
\href{https://doi.org//10.1103/PhysRevD.96.036019}{Phys. Rev. D \textbf{96}, 036019 (2017)},
\href{https://arxiv.org/abs/1707.03972}{\color{teal}{[arXiv:1707.03972 [hep-ph]]}}.

\bibitem{Polyakov:2019lbq}
M.~V.~Polyakov and B.~D.~Sun,
Gravitational form factors of a spin one particle,
\href{https://doi.org//10.1103/PhysRevD.100.036003}{Phys. Rev. D \textbf{100}, 036003 (2019)},
\href{https://arxiv.org/abs/1903.02738}{\color{teal}{[arXiv:1903.02738 [hep-ph]]}}.

\bibitem{Sun:2020wfo}
B.~D.~Sun and Y.~B.~Dong,
Gravitational form factors of $\rho$ meson with a light-cone constituent quark model,
\href{https://doi.org//10.1103/PhysRevD.101.096008}{Phys. Rev. D \textbf{101}, 096008 (2020)},
\href{https://arxiv.org/abs/2002.02648}{\color{teal}{[arXiv:2002.02648 [hep-ph]]}}.

\bibitem{ParticleDataGroup:2024cfk}
S.~Navas \textit{et al.} [Particle Data Group],
Review of particle physics,
\href{https://doi.org//10.1103/PhysRevD.110.030001}{Phys. Rev. D \textbf{110}, 030001 (2024)}.


\bibitem{LopezCastro:2000cv}
G.~Lopez Castro and A.~Mariano,
Determination of the Delta++ magnetic dipole moment,
\href{https://doi.org//10.1016/S0370-2693(01)00980-7}{Phys. Lett. B \textbf{517}, 339-344 (2001)},
\href{https://arxiv.org/abs/nucl-th/0006031}{\color{teal}{[arXiv:nucl-th/0006031 [nucl-th]]}}.

\bibitem{Kotulla:2002cg}
M.~Kotulla, J.~Ahrens, J.~R.~M.~Annand, R.~Beck, G.~Caselotti, L.~S.~Fog, D.~Hornidge, S.~Janssen, B.~Krusche and J.~C.~McGeorge, \textit{et al.}
The Reaction gamma p ---{\ensuremath{>}} pi zero gamma-prime p and the magnetic dipole moment of the delta+(1232) resonance,
\href{https://doi.org//10.1103/PhysRevLett.89.272001}{Phys. Rev. Lett. \textbf{89}, 272001 (2002)},
\href{https://arxiv.org/abs/nucl-ex/0210040}{\color{teal}{[arXiv:nucl-ex/0210040 [nucl-ex]]}}.


\bibitem{Alexandrou:2008bn}
C.~Alexandrou, T.~Korzec, G.~Koutsou, T.~Leontiou, C.~Lorce, J.~W.~Negele, V.~Pascalutsa, A.~Tsapalis and M.~Vanderhaeghen,
Delta-baryon electromagnetic form factors in lattice QCD,
\href{https://doi.org//10.1103/PhysRevD.79.014507}{Phys. Rev. D \textbf{79}, 014507 (2009)},
\href{https://arxiv.org/abs/0810.3976}{\color{teal}{[arXiv:0810.3976 [hep-lat]]}}.

\bibitem{Boinepalli:2009sq}
S.~Boinepalli, D.~B.~Leinweber, P.~J.~Moran, A.~G.~Williams, J.~M.~Zanotti and J.~B.~Zhang,
Precision electromagnetic structure of decuplet baryons in the chiral regime,
\href{https://doi.org//10.1103/PhysRevD.80.054505}{Phys. Rev. D \textbf{80}, 054505 (2009)},
\href{https://arxiv.org/abs/0902.4046}{\color{teal}{[arXiv:0902.4046 [hep-lat]]}}.

\bibitem{Leinweber:1992hy}
D.~B.~Leinweber, T.~Draper and R.~M.~Woloshyn,
Decuplet baryon structure from lattice QCD,
\href{https://doi.org//10.1103/PhysRevD.46.3067}{Phys. Rev. D \textbf{46}, 3067-3085 (1992)},
\href{https://arxiv.org/abs/hep-lat/9208025}{\color{teal}{[arXiv:hep-lat/9208025 [hep-lat]]}}.

\bibitem{Aubin:2008qp}
C.~Aubin, K.~Orginos, V.~Pascalutsa and M.~Vanderhaeghen,
Magnetic Moments of Delta and Omega- Baryons with Dynamical Clover Fermions,
\href{https://doi.org//10.1103/PhysRevD.79.051502}{Phys. Rev. D \textbf{79}, 051502 (2009)},
\href{https://arxiv.org/abs/0811.2440 }{\color{teal}{[arXiv:0811.2440 [hep-lat]]}}.

\bibitem{Seth:2014dxa}
K.~K.~Seth,
First Measurements of Form Factors of Pion, Kaon, Proton, and Hyperons for the Highest Timelike Momentum Transfers,
\href{https://doi.org//10.1088/1742-6596/556/1/012009}{J. Phys. Conf. Ser. \textbf{556}, 012009 (2014)}.

\bibitem{Junker:2019vvy}
O.~Junker, S.~Leupold, E.~Perotti and T.~Vitos,
Electromagnetic form factors of the transition from the spin-3/2 $\Sigma$ to the $\Lambda$ hyperon,
\href{https://doi.org//10.1103/PhysRevC.101.015206}{Phys. Rev. C \textbf{101}, 015206 (2020)},
\href{https://arxiv.org/abs/1910.07396}{\color{teal}{[arXiv:1910.07396 [hep-ph]]}}.

\bibitem{Dai:2021yqr}
A.~X.~Dai, Z.~Y.~Li, L.~Chang and J.~J.~Xie,
Electromagnetic form factors of neutron and neutral hyperons in the oscillating point of view*,
\href{https://doi.org//10.1088/1674-1137/ac5f9c}{Chin. Phys. C \textbf{46}, 073104 (2022)},
\href{https://arxiv.org/abs/2112.06264 }{\color{teal}{[arXiv:2112.06264 [hep-ph]]}}.

\bibitem{Lin:2008rb}
H.~W.~Lin,
Hyperon Physics from Lattice QCD,
\href{https://doi.org//10.1016/j.nuclphysbps.2009.01.029}{Nucl. Phys. B Proc. Suppl. \textbf{187}, 200-207 (2009)},
\href{https://arxiv.org/abs/0812.0411}{\color{teal}{[arXiv:0812.0411 [hep-lat]]}}.

\cite{Bartel:1968tw}
\bibitem{Bartel:1968tw}
W.~Bartel, B.~Dudelzak, H.~Krehbiel, J.~McElroy, U.~Meyer-Berkhout, W.~Schmidt, V.~Walther and G.~Weber,
Electroproduction of pions near the $H(1236)$ isobar and the form-factor $G^*_M(q^2)$ of the $({\gamma} NH)$ vertex,
\href{https://doi.org//10.1016/0370-2693(68)90155-X}{Phys. Lett. B \textbf{28}, 148-151 (1968)}.

\bibitem{Stein:1975yy}
S.~Stein, W.~B.~Atwood, E.~D.~Bloom, R.~L.~Cottrell, H.~C.~DeStaebler, C.~L.~Jordan, H.~Piel, C.~Y.~Prescott, R.~Siemann and R.~E.~Taylor,
Electron Scattering at 4-Degrees with Energies of 4.5-GeV - 20-GeV,
\href{https://doi.org//10.1103/PhysRevD.12.1884}{Phys. Rev. D \textbf{12}, 1884 (1975)}.


\bibitem{Frolov:1998pw}
V.~V.~Frolov, G.~S.~Adams, A.~Ahmidouch, C.~S.~Armstrong, K.~Assamagan, S.~Avery, O.~K.~Baker, P.~E.~Bosted, V.~Burkert and R.~Carlini, \textit{et al.}
Electroproduction of the Delta (1232) resonance at high momentum transfer,
\href{https://doi.org//10.1103/PhysRevLett.82.45}{Phys. Rev. Lett. \textbf{82}, 45-48 (1999)},
\href{https://arxiv.org/abs/hep-ex/9808024}{\color{teal}{[arXiv:hep-ex/9808024 [hep-ex]]}}.

\bibitem{Caia:2004pm}
G.~L.~Caia, V.~Pascalutsa, J.~A.~Tjon and L.~E.~Wright,
gamma* N Delta form-factors from a relativistic dynamical model of pion electroproduction,
\href{https://doi.org//10.1103/PhysRevC.70.032201}{Phys. Rev. C \textbf{70}, 032201 (2004)},
\href{https://arxiv.org/abs/nucl-th/0407069}{\color{teal}{[arXiv:nucl-th/0407069 [nucl-th]]}}.

\bibitem{CLAS:2001cbm}
K.~Joo \textit{et al.} [CLAS],
Q**2 dependence of quadrupole strength in the gamma* p ---{\ensuremath{>}} Delta+(1232) ---{\ensuremath{>}} p pi0 transition,
\href{https://doi.org//10.1103/PhysRevLett.88.122001}{Phys. Rev. Lett. \textbf{88}, 122001 (2002)},
\href{https://arxiv.org/abs/hep-ex/0110007 }{\color{teal}{[arXiv:hep-ex/0110007 [hep-ex]]}}.
Phys. Rev. Lett. \textbf{88}, 122001 (2002)
doi:10.1103/PhysRevLett.88.122001
[arXiv:hep-ex/0110007 [hep-ex]].

\bibitem{CLAS:2003hro}
A.~S.~Biselli \textit{et al.} [CLAS],
Study of e p ---{\ensuremath{>}} e p pi0 in the Delta(1232) mass region using polarization asymmetries,
\href{https://doi.org//10.1103/PhysRevC.68.035202}{Phys. Rev. C \textbf{68}, 035202 (2003)},
\href{https://arxiv.org/abs/nucl-ex/0307004}{\color{teal}{[arXiv:nucl-ex/0307004 [nucl-ex]]}}.

\bibitem{CLAS:2009ces}
I.~G.~Aznauryan \textit{et al.} [CLAS],
Electroexcitation of nucleon resonances from CLAS data on single pion electroproduction,
\href{https://doi.org//10.1103/PhysRevC.80.055203}{Phys. Rev. C \textbf{80}, 055203 (2009)},
\href{https://arxiv.org/abs/0909.2349 }{\color{teal}{[arXiv:0909.2349 [nucl-ex]]}}.


\bibitem{Guo:2007dw}
L.~Guo, D.~P.~Weygand, M.~Battaglieri, R.~D.~Vita, V.~Kubarovsky, P.~Stoler, M.~J.~Amaryan, P.~Ambrozewicz, M.~Anghinolfi and G.~Asryan, \textit{et al.}
Cascade production in the reactions gamma p ---{\ensuremath{>}} K+ K+ (X) and gamma p ---{\ensuremath{>}} K+ K+ pi- (X),
\href{https://doi.org//10.1103/PhysRevC.76.025208}{Phys. Rev. C \textbf{76}, 025208 (2007)},
\href{https://arxiv.org/abs/nucl-ex/0702027 }{\color{teal}{[arXiv:nucl-ex/0702027 [nucl-ex]]}}.

\bibitem{CLAS:2018kvn}
J.~T.~Goetz \textit{et al.} [CLAS],
Study of $\Xi^*$ Photoproduction from Threshold to $W = 3.3$ GeV,
\href{https://doi.org//10.1103/PhysRevC.98.062201}{Phys. Rev. C \textbf{98}, 062201 (2018)},
\href{https://arxiv.org/abs/1809.00074}{\color{teal}{[arXiv:1809.00074 [nucl-ex]]}}.

\bibitem{Dobbs:2014ifa}
S.~Dobbs, A.~Tomaradze, T.~Xiao, K.~K.~Seth and G.~Bonvicini,
First measurements of timelike form factors of the hyperons, $\Lambda^0$,  $\Sigma^0$,  $\Sigma^+$,  $\Xi^0$,  $\Xi^-$,  and $\Omega^-$, and evidence of diquark correlations,
\href{https://doi.org//10.1016/j.physletb.2014.10.025}{Phys. Lett. B \textbf{739}, 90-94 (2014)},
\href{https://arxiv.org/abs/1410.8356}{\color{teal}{[arXiv:1410.8356 [hep-ex]]}}.

\bibitem{BESIII:2022kzc}
M.~Ablikim \textit{et al.} [BESIII],
Study of $e^+e^-\to\Omega^-\bar{\Omega}^+$ at center-of-mass energies from 3.49 to 3.67~GeV,
\href{https://doi.org//10.1103/PhysRevD.107.052003}{Phys. Rev. D \textbf{107}, 052003 (2023)},
\href{https://arxiv.org/abs/2212.03693}{\color{teal}{[arXiv:2212.03693 [hep-ex]]}}.

\bibitem{Alexandrou:2010jv}
C.~Alexandrou, T.~Korzec, G.~Koutsou, J.~W.~Negele and Y.~Proestos,
The Electromagnetic form factors of the $\Omega^-$ in lattice QCD,
\href{https://doi.org//10.1103/PhysRevD.82.034504}{Phys. Rev. D \textbf{82}, 034504 (2010)},
\href{https://arxiv.org/abs/1006.0558}{\color{teal}{[arXiv:1006.0558 [hep-lat]]}}.



\bibitem{Beck:1997ew}
R.~Beck, H.~P.~Krahn, J.~Ahrens, H.~J.~Arends, G.~Audit, A.~Braghieri, N.~d'Hose, S.~J.~Hall, V.~Isbert and J.~D.~Kellie, \textit{et al.}
Measurement of the E2/M1 ratio in the N ---{\ensuremath{>}} Delta transition using the reaction p(gamma(pol.),p)pi0,
\href{https://doi.org//10.1103/PhysRevLett.78.606}{Phys. Rev. Lett. \textbf{78}, 606-609 (1997)}.

\bibitem{Pascalutsa:2005ts}
V.~Pascalutsa and M.~Vanderhaeghen,
Electromagnetic nucleon-to-Delta transition in chiral effective-field theory,
\href{https://doi.org//10.1103/PhysRevLett.95.232001}{Phys. Rev. Lett. \textbf{95}, 232001 (2005)},
\href{https://arxiv.org/abs/hep-ph/0508060}{\color{teal}{[arXiv:hep-ph/0508060 [hep-ph]]}}.

\bibitem{A1:2008ocu}
S.~Stave \textit{et al.} [A1],
Measurements of the gamma* p ---{\ensuremath{>}} Delta Reaction At Low Q**2: Probing the Mesonic Contribution,
\href{https://doi.org//10.1103/PhysRevC.78.025209}{Phys. Rev. C \textbf{78}, 025209 (2008)},
\href{https://arxiv.org/abs/0803.2476}{\color{teal}{[arXiv:0803.2476 [hep-ex]]}}.

\bibitem{Sparveris:2013ena}
N.~Sparveris, S.~Stave, P.~Achenbach, C.~Ayerbe Gayoso, D.~Baumann, J.~Bernauer, A.~M.~Bernstein, R.~Bohm, D.~Bosnar and T.~Botto, \textit{et al.}
Measurements of the $\gamma$*p $\rightarrow$  $\Delta$ reaction at low Q$^{2}$,
\href{https://doi.org//10.1140/epja/i2013-13136-2}{Eur. Phys. J. A \textbf{49}, 136 (2013)},
\href{https://arxiv.org/abs/1307.0751 }{\color{teal}{[arXiv:1307.0751 [nucl-ex]]}}.

\bibitem{Blanpied:1997zz}
G.~Blanpied, M.~Blecher, A.~Caracappa, C.~Djalali, G.~Giordano, K.~Hicks, S.~Hoblit, M.~Khandaker, O.~C.~Kistner and A.~Kuczewski, \textit{et al.}
N --{\ensuremath{>}} Delta Transition from Simultaneous Measurements of p (gamma --{\ensuremath{>}}, pi) and p (gamma --{\ensuremath{>}}, gamma),
\href{https://doi.org/10.1103/PhysRevLett.79.4337}{Phys. Rev. Lett. \textbf{79}, 4337-4340 (1997)}.

\bibitem{Blanpied:2001ae}
G.~Blanpied, M.~Blecher, A.~Caracappa, R.~Deininger, C.~Djalali, G.~Giordano, K.~Hicks, S.~Hoblit, M.~Khandaker and O.~C.~Kistner, \textit{et al.}
N ---{\ensuremath{>}} delta transition and proton polarizabilities from measurements of p (gamma polarized, gamma), p (gamma polarized, pi0), and p (gamma polarized, pi+),
\href{https://doi.org//10.1103/PhysRevC.64.025203}{Phys. Rev. C \textbf{64}, 025203 (2001)}.

\bibitem{Lee:1997jk}
F.~X.~Lee,
Determination of decuplet baryon magnetic moments from QCD sum rules,
\href{https://doi.org//10.1103/PhysRevD.57.1801}{Phys. Rev. D \textbf{57}, 1801-1821 (1998)},
\href{https://arxiv.org/abs/hep-ph/9708323}{\color{teal}{[arXiv:hep-ph/9708323 [hep-ph]]}}.

\bibitem{Aliev:2009pd}
T.~M.~Aliev, K.~Azizi and M.~Savci,
Electric Quadrupole and Magnetic Octupole Moments of the Light Decuplet Baryons Within Light Cone QCD Sum Rules,
\href{https://doi.org//10.1016/j.physletb.2009.10.026}{Phys. Lett. B \textbf{681}, 240-246 (2009)},
\href{https://arxiv.org/abs/0904.2485}{\color{teal}{[arXiv:0904.2485 [hep-ph]]}}.

\bibitem{Azizi:2009egn}
K.~Azizi,
Magnetic Dipole, Electric Quadrupole and Magnetic Octupole Moments of the Delta Baryons in Light Cone QCD Sum Rules,
\href{https://doi.org//10.1140/epjc/s10052-009-0988-0}{Eur. Phys. J. C \textbf{61}, 311-319 (2009)},
\href{https://arxiv.org/abs/0811.2670 }{\color{teal}{[arXiv:0811.2670 [hep-ph]]}}.

\bibitem{Aliev:2015qea}
T.~M.~Aliev, K.~Azizi, T.~Barakat and M.~Savc{\i},
Diagonal and transition magnetic moments of negative parity heavy baryons in QCD sum rules,
\href{https://doi.org//10.1103/PhysRevD.92.036004}{Phys. Rev. D \textbf{92}, 036004 (2015)},
\href{https://arxiv.org/abs/1505.07977 }{\color{teal}{[arXiv:1505.07977 [hep-ph]]}}.

\bibitem{Buchmann:2002et}
A.~J.~Buchmann and R.~F.~Lebed,
Baryon charge radii and quadrupole moments in the 1/N(c) expansion: The three flavor case,
\href{https://doi.org//10.1103/PhysRevD.67.016002}{Phys. Rev. D \textbf{67}, 016002 (2003)},
\href{https://arxiv.org/abs/hep-ph/0207358}{\color{teal}{[arXiv:hep-ph/0207358 [hep-ph]]}}.

\bibitem{Flores-Mendieta:2015wir}
R.~Flores-Mendieta and M.~A.~Rivera-Ruiz,
Dirac form factors and electric charge radii of baryons in the combined chiral and 1/N$_c$ expansions,
\href{https://doi.org//10.1103/PhysRevD.92.094026}{Phys. Rev. D \textbf{92}, 094026 (2015)},
\href{https://arxiv.org/abs/1511.02932}{\color{teal}{[arXiv:1511.02932 [hep-ph]]}}.

\bibitem{BandaGuzman:2020fxx}
V.~M.~Banda Guzm{\'a}n, R.~Flores-Mendieta, J.~Hern{\'a}ndez and F.~d.~Rosales-Aldape,
Baryon quadrupole moment in the 1/$N_c$ expansion of QCD,
\href{https://doi.org//10.1103/PhysRevD.101.074018}{Phys. Rev. D \textbf{101}, 074018 (2020)},
\href{https://arxiv.org/abs/2002.09256}{\color{teal}{[arXiv:2002.09256 [hep-ph]]}}.

\bibitem{GonzalezdeUrreta:2012zz}
E.~J.~Gonzalez de Urreta and N.~N.~Scoccola,
Consistent analysis of the masses and decays of the [70, 1-] baryons in the 1/N(c) expansion,
\href{https://doi.org//10.1063/1.3701227}{AIP Conf. Proc. \textbf{1432}, 265-268 (2012)}.

\bibitem{Matagne:2012zz}
N.~Matagne and F.~Stancu,
Excited baryons in the 1/N(c) expansion,
\href{https://doi.org//10.1063/1.3701197}{AIP Conf. Proc. \textbf{1432}, 106-111 (2012)}.

\bibitem{Goity:2010zz}
J.~L.~Goity, C.~P.~Jayalath and N.~N.~Scoccola,
The 1/N(c) expansion in baryons,
\href{https://doi.org//10.1063/1.3483316}{AIP Conf. Proc. \textbf{1257}, 181-188 (2010)}.

\bibitem{Jayalath:2010zz}
C.~P.~Jayalath, J.~L.~Goity and N.~N.~Scoccola,
Negative parity baryon decays in the 1/N(c) expansion,
\href{https://doi.org//10.1063/1.3483391}{AIP Conf. Proc. \textbf{1257}, 548-552 (2010)}.

\bibitem{Kim:2019gka}
J.~Y.~Kim and H.~C.~Kim,
Electromagnetic form factors of the baryon decuplet with flavor SU(3) symmetry breaking,
\href{https://doi.org//10.1140/epjc/s10052-019-7079-7}{Eur. Phys. J. C \textbf{79}, 570 (2019)},
\href{https://arxiv.org/abs/1905.04017}{\color{teal}{[arXiv:1905.04017 [hep-ph]]}}.

\bibitem{Geng:2009ys}
L.~S.~Geng, J.~Martin Camalich and M.~J.~Vicente Vacas,
Electromagnetic structure of the lowest-lying decuplet resonances in covariant chiral perturbation theory,
\href{https://doi.org//10.1103/PhysRevD.80.034027}{Phys. Rev. D \textbf{80}, 034027 (2009)},
\href{https://arxiv.org/abs/0907.0631}{\color{teal}{[arXiv:0907.0631 [hep-ph]]}}.

\bibitem{Berger:2004yi}
K.~Berger, R.~F.~Wagenbrunn and W.~Plessas,
Covariant baryon charge radii and magnetic moments in a chiral constituent quark model,
\href{https://doi.org//10.1103/PhysRevD.70.094027}{Phys. Rev. D \textbf{70}, 094027 (2004)},
\href{https://arxiv.org/abs/nucl-th/0407009}{\color{teal}{[arXiv:nucl-th/0407009 [nucl-th]]}}.

\bibitem{Oh:1995hn}
Y.~s.~Oh,
Electric quadrupole moments of the decuplet baryons in the Skyrme model,
\href{https://doi.org//10.1142/S0217732395001137}{Mod. Phys. Lett. A \textbf{10}, 1027-1034 (1995)},
\href{https://arxiv.org/abs/hep-ph/9506308}{\color{teal}{[arXiv:hep-ph/9506308 [hep-ph]]}}.

\bibitem{Krivoruchenko:1991pm}
M.~I.~Krivoruchenko and M.~M.~Giannini,
Quadrupole moments of the decuplet baryons,
\href{https://doi.org//10.1103/PhysRevD.43.3763}{Phys. Rev. D \textbf{43}, 3763-3765 (1991)}.

\bibitem{Wang:2023bjp}
J.~Wang, D.~Fu and Y.~Dong,
Form factors of decuplet baryons in a covariant quark{\textendash}diquark approach,
\href{https://doi.org/10.1140/epjc/s10052-024-12406-4}{Eur. Phys. J. C \textbf{84}, 79 (2024)},
\href{https://arxiv.org/abs/2311.07149}{\color{teal}{[arXiv:2311.07149 [hep-ph]]}}.


\bibitem{Diehl:2003ny}
M.~Diehl,
Generalized parton distributions,
\href{https://doi.org//10.1016/j.physrep.2003.08.002}{Phys. Rept. \textbf{388}, 41-277 (2003)},
\href{https://arxiv.org/abs/hep-ph/0307382}{\color{teal}{[arXiv:hep-ph/0307382 [hep-ph]]}}.



\bibitem{Kumano:2017lhr}
S.~Kumano, Q.~T.~Song and O.~V.~Teryaev,
Hadron tomography by generalized distribution amplitudes in pion-pair production process $\gamma^* \gamma \rightarrow \pi^0 \pi^0 $ and gravitational form factors for pion,
\href{https://doi.org//10.1103/PhysRevD.97.014020}{Phys. Rev. D \textbf{97}, 014020 (2018)},
\href{https://arxiv.org/abs/1711.08088}{\color{teal}{[arXiv:1711.08088 [hep-ph]]}}.

\bibitem{Fu:2022rkn}
D.~Fu, B.~D.~Sun and Y.~Dong,
Electromagnetic and gravitational form factors of {\ensuremath{H}} resonance in a covariant quark-diquark approach,
\href{https://doi.org//10.1103/PhysRevD.105.096002}{Phys. Rev. D \textbf{105}, 096002 (2022)},
\href{https://arxiv.org/abs/2201.08059}{\color{teal}{[arXiv:2201.08059 [hep-ph]]}}.

\bibitem{Kim:2020lrs}
J.~Y.~Kim and B.~D.~Sun,
Gravitational form factors of a baryon with spin-3/2,
\href{https://doi.org//10.1140/epjc/s10052-021-08852-z}{Eur. Phys. J. C \textbf{81}, 85 (2021)},
\href{https://arxiv.org/abs/2011.00292}{\color{teal}{[arXiv:2011.00292 [hep-ph]]}}.

\bibitem{Pefkou:2021fni}
D.~A.~Pefkou, D.~C.~Hackett and P.~E.~Shanahan,
Gluon gravitational structure of hadrons of different spin,
\href{https://doi.org//10.1103/PhysRevD.105.054509}{Phys. Rev. D \textbf{105}, 054509 (2022)},
\href{https://arxiv.org/abs/2107.10368}{\color{teal}{[arXiv:2107.10368 [hep-lat]]}}.

\bibitem{Panteleeva:2020ejw}
J.~Y.~Panteleeva and M.~V.~Polyakov,
Quadrupole pressure and shear forces inside baryons in the large $N_c$ limit,
\href{https://doi.org//10.1016/j.physletb.2020.135707}{Phys. Lett. B \textbf{809}, 135707 (2020)},
\href{https://arxiv.org/abs/2004.02912 }{\color{teal}{[arXiv:2004.02912 [hep-ph]]}}.

\bibitem{Dehghan:2023ytx}
Z.~Dehghan, K.~Azizi and U.~{\"O}zdem,
Gravitational form factors of the {\ensuremath{H}} baryon via QCD sum rules,
\href{https://doi.org//10.1103/PhysRevD.108.094037}{Phys. Rev. D \textbf{108}, 094037 (2023)},
\href{https://arxiv.org/abs/2307.14880}{\color{teal}{[arXiv:2307.14880 [hep-ph]]}}.

\bibitem{Dehghan:2025eov}
Z.~Dehghan and K.~Azizi,
Mechanical properties of the {\ensuremath{\Omega}}- baryon from gravitational form factors,
\href{https://doi.org//10.1103/x6f5-cbcc}{Phys. Rev. D \textbf{112}, 054014 (2025)},
\href{https://arxiv.org/abs/2507.14840}{\color{teal}{[arXiv:2507.14840 [hep-ph]]}}.

\bibitem{Perevalova:2016dln}
I.~A.~Perevalova, M.~V.~Polyakov and P.~Schweitzer,
On LHCb pentaquarks as a baryon-$\psi$(2S) bound state: prediction of isospin-$\frac3{2}$ pentaquarks with hidden charm,
\href{https://doi.org/10.1103/PhysRevD.94.054024}{Phys. Rev. D \textbf{94}, 054024 (2016)},
\href{https://arxiv.org/abs/1607.07008}{\color{teal}{[arXiv:1607.07008 [hep-ph]]}}.

\bibitem{Cotogno:2019vjb}
S.~Cotogno, C.~Lorc{\'e}, P.~Lowdon and M.~Morales,
Covariant multipole expansion of local currents for massive states of any spin,
\href{https://doi.org/10.1103/PhysRevD.101.056016}{Phys. Rev. D \textbf{101}, 056016 (2020)},
\href{https://arxiv.org/abs/1912.08749}{\color{teal}{[arXiv:1912.08749 [hep-ph]]}}.

\bibitem{Kim:2021zbz}
J.~Y.~Kim, H.~C.~Kim and M.~V.~Polyakov,
Light-cone distribution amplitudes of the nucleon and {\ensuremath{H}} baryon,
\href{https://doi.org/10.1007/JHEP11(2021)039}{JHEP \textbf{11}, 039 (2021)},
\href{https://arxiv.org/abs/2110.05889}{\color{teal}{[arXiv:2110.05889 [hep-ph]]}}.


\bibitem{Alharazin:2022wjj}
H.~Alharazin, E.~Epelbaum, J.~Gegelia, U.~G.~Mei{\ss}ner and B.~D.~Sun,
Gravitational form factors of the delta resonance in chiral EFT,
\href{https://doi.org/10.1140/epjc/s10052-022-10882-0}{Eur. Phys. J. C \textbf{82}, 907 (2022)},
\href{https://arxiv.org/abs/2209.01233}{\color{teal}{[arXiv:2209.01233 [hep-ph]]}}.

\bibitem{Alharazin:2022xvp}
H.~Alharazin, B.~D.~Sun, E.~Epelbaum, J.~Gegelia and U.~G.~Mei{\ss}ner,
Local spatial densities for composite spin-3/2 systems,
\href{https://doi.org/10.1007/JHEP02(2023)163}{JHEP \textbf{02}, 163 (2023)},
\href{https://arxiv.org/abs/2212.11505}{\color{teal}{[arXiv:2212.11505 [hep-ph]]}}.

\bibitem{Fu:2023ijy}
D.~Fu, J.~Wang and Y.~Dong,
Form factors of {\ensuremath{\Omega}}- in a covariant quark-diquark approach,
\href{https://doi.org/10.1103/PhysRevD.108.076023}{Phys. Rev. D \textbf{108}, 076023 (2023)},
\href{https://arxiv.org/abs/2306.04869}{\color{teal}{[arXiv:2306.04869 [hep-ph]]}}.


\bibitem{Kim:2022bwn}
J.~Y.~Kim,
Parametrization of transition energy-momentum tensor form factors,
\href{https://doi.org/10.1016/j.physletb.2022.137442}{Phys. Lett. B \textbf{834}, 137442 (2022)},
\href{https://arxiv.org/abs/2206.10202}{\color{teal}{[arXiv:2206.10202 [hep-ph]]}}.

\bibitem{Ozdem:2022zig}
U.~{\"O}zdem and K.~Azizi,
Gravitational transition form factors of N {\textrightarrow} {\ensuremath{H}} via QCD light-cone sum rules,
\href{https://doi.org/10.1007/JHEP03(2023)048}{JHEP \textbf{03}, 048 (2023)},
\href{https://arxiv.org/abs/2212.07290}{\color{teal}{[arXiv:2212.07290 [hep-ph]]}}.

\bibitem{Kim:2023yhp}
J.~Y.~Kim,
Quark distribution functions and spin-flavor structures in N{\textrightarrow}{\ensuremath{H}} transitions,
\href{https://doi.org/10.1103/PhysRevD.108.034024}{Phys. Rev. D \textbf{108}, 034024 (2023)},
\href{https://arxiv.org/abs/2305.12714}{\color{teal}{[arXiv:2305.12714 [hep-ph]]}}.

\bibitem{Alharazin:2023zzc}
H.~Alharazin, B.~D.~Sun, E.~Epelbaum, J.~Gegelia and U.~G.~Mei{\ss}ner,
Gravitational~$p \to H^+ $ transition form factors in chiral perturbation theory,
\href{https://doi.org/10.1007/JHEP03(2024)007}{JHEP \textbf{03}, 007 (2024)},
\href{https://arxiv.org/abs/2312.05193}{\color{teal}{[arXiv:2312.05193 [hep-ph]]}}.

\bibitem{Goharipour:2024atx}
M.~Goharipour \textit{et al.} [MMGPDs],
Impact of JLab data on the determination of GPDs at zero skewness and
new insights from transition form factors $N\rightarrow H$,
\href{https://doi.org/10.1103/PhysRevD.109.074042}{Phys. Rev. D \textbf{109}, 074042 (2024)},
\href{https://arxiv.org/abs/2403.19384}{\color{teal}{[arXiv:2403.19384 [hep-ph]]}}.

\bibitem{Ralston:1979ys}
J.~P.~Ralston and D.~E.~Soper,
Production of Dimuons from High-Energy Polarized Proton Proton Collisions,
\href{https://doi.org//10.1016/0550-3213(79)90082-8}{Nucl. Phys. B \textbf{152}, 109 (1979)}.

\bibitem{Jaffe:1991kp}
R.~L.~Jaffe and X.~D.~Ji,
Chiral odd parton distributions and polarized Drell-Yan,
\href{https://doi.org//10.1103/PhysRevLett.67.552}{Phys. Rev. Lett. \textbf{67}, 552-555 (1991)}.

\bibitem{Jaffe:1991ra}
R.~L.~Jaffe and X.~D.~Ji,
Chiral odd parton distributions and Drell-Yan processes,
\href{https://doi.org//10.1016/0550-3213(92)90110-W}{Nucl. Phys. B \textbf{375}, 527-560 (1992)}.

\bibitem{Barone:2001sp}
V.~Barone, A.~Drago and P.~G.~Ratcliffe,
Transverse polarisation of quarks in hadrons,
\href{https://doi.org//10.1016/S0370-1573(01)00051-5}{Phys. Rept. \textbf{359}, 1-168 (2002)},
\href{https://arxiv.org/abs/hep-ph/0104283}{\color{teal}{[arXiv:hep-ph/0104283 [hep-ph]]}}.

\bibitem{He:1994gz}
H.~x.~He and X.~D.~Ji,
The Nucleon's tensor charge,
\href{https://doi.org//10.1103/PhysRevD.52.2960}{Phys. Rev. D \textbf{52}, 2960-2963 (1995)},
\href{https://arxiv.org/abs/hep-ph/9412235}{\color{teal}{[arXiv:hep-ph/9412235 [hep-ph]]}}.

\bibitem{Pasquini:2005dk}
B.~Pasquini, M.~Pincetti and S.~Boffi,
Chiral-odd generalized parton distributions in constituent quark models,
\href{https://doi.org//10.1103/PhysRevD.72.094029}{Phys. Rev. D \textbf{72}, 094029 (2005)},
\href{https://arxiv.org/abs/hep-ph/0510376}{\color{teal}{[arXiv:hep-ph/0510376 [hep-ph]]}}.

\bibitem{Gamberg:2001qc}
L.~P.~Gamberg and G.~R.~Goldstein,
Flavor spin symmetry estimate of the nucleon tensor charge,
\href{https://doi.org//10.1103/PhysRevLett.87.242001}{Phys. Rev. Lett. \textbf{87}, 242001 (2001)},
\href{https://arxiv.org/abs/hep-ph/0107176}{\color{teal}{[arXiv:hep-ph/0107176 [hep-ph]]}}.

\bibitem{kucukarslan:2016xhx}
A.~kucukarslan, U.~Ozdem and A.~Ozpineci,
Tensor form factors of the octet hyperons in QCD,
\href{https://doi.org//10.1103/PhysRevD.94.094010}{Phys. Rev. D \textbf{94}, 094010 (2016)},
\href{https://arxiv.org/abs/1610.08358}{\color{teal}{[arXiv:1610.08358 [hep-ph]]}}.


\bibitem{Gutsche:2016xff}
T.~Gutsche, M.~A.~Ivanov, J.~G.~Korner, S.~Kovalenko and V.~E.~Lyubovitskij,
Nucleon tensor form factors in a relativistic confined quark model,
\href{https://doi.org//10.1103/PhysRevD.94.114030}{Phys. Rev. D \textbf{94}, 114030 (2016)},
\href{https://arxiv.org/abs/1608.00420}{\color{teal}{[arXiv:1608.00420 [hep-ph]]}}.

\bibitem{Azizi:2019ytx}
K.~Azizi and U.~{\"O}zdem,
Nucleon{\textquoteright}s energy{\textendash}momentum tensor form factors in light-cone QCD,
\href{https://doi.org//10.1140/epjc/s10052-020-7676-5}{Eur. Phys. J. C \textbf{80}, 104 (2020)},
\href{https://arxiv.org/abs/1908.06143}{\color{teal}{[arXiv:1908.06143 [hep-ph]]}}.

\bibitem{Ozdem:2020vpt}
U.~{\"O}zdem,
Isovector and isoscalar tensor form factors of $N(1535) \rightarrow N$ transition in light-cone QCD,
\href{https://doi.org//10.1103/PhysRevD.102.014001}{Phys. Rev. D \textbf{102}, 014001 (2020)},
\href{https://arxiv.org/abs/2004.12312}{\color{teal}{[arXiv:2004.12312 [hep-ph]]}}.

\bibitem{Ozdem:2021zbn}
U.~{\"O}zdem,
Tensor form factors of N(1535) state via light-cone QCD,
\href{https://doi.org//10.1016/j.cjph.2021.04.018}{Chin. J. Phys. \textbf{72}, 93-99 (2021)}.

\bibitem{Fu:2024kfx}
D.~Fu, Y.~Dong and S.~Kumano,
Transversity generalized parton distributions in spin-3/2 particles,
\href{https://doi.org//10.1103/PhysRevD.109.096006}{Phys. Rev. D \textbf{109}, 096006 (2024)},
\href{https://arxiv.org/abs/2402.11561}{\color{teal}{[arXiv:2402.11561 [hep-ph]]}}.

\bibitem{Fu:2025upc}
D.~Fu, Y.~Dong and S.~Kumano,
Transversity generalized parton distributions of the {\ensuremath{\Delta}} with the diquark spectator model,
\href{https://doi.org//10.1103/d17w-jkr8}{Phys. Rev. D \textbf{112}, 096027 (2025)},
\href{https://arxiv.org/abs/2508.15245}{\color{teal}{[arXiv:2508.15245 [hep-ph]]}}.

\bibitem{Asmaee:2025elo}
Z.~Asmaee, N.~Hajirasouliha and K.~Azizi,
Tensor form factors of the {\ensuremath{\Delta^+}} baryon induced by isovector and isoscalar currents in QCD,
\href{https://doi.org//10.1103/mlt1-gvnw}{Phys. Rev. D \textbf{113}, 054002 (2026)},
\href{https://arxiv.org/abs/2511.12360}{\color{teal}{[arXiv:2511.12360 [hep-ph]]}}.

\bibitem{Aliev:2016jnp}
T.~M.~Aliev, K.~Azizi and H.~Sundu,
Radial Excitations of the Decuplet Baryons,
\href{https://doi.org//10.1140/epjc/s10052-017-4782-0}{Eur. Phys. J. C \textbf{77}, 222 (2017)},
\href{https://arxiv.org/abs/1612.03661}{\color{teal}{[arXiv:1612.03661 [hep-ph]]}}.

\bibitem{Aliev:2010dw}
T.~M.~Aliev, K.~Azizi and M.~Savci,
Strong Coupling Constants of Decuplet Baryons with Vector Mesons,
\href{https://doi.org//10.1103/PhysRevD.82.096006}{Phys. Rev. D \textbf{82}, 096006 (2010)},
\href{https://arxiv.org/abs/1007.3389}{\color{teal}{[arXiv:1007.3389 [hep-ph]]}}.

\bibitem{Azizi:2016ddw}
K.~Azizi and G.~Bozk{\i}r,
Decuplet baryons in a hot medium,
\href{https://doi.org/10.1140/epjc/s10052-016-4370-8}{Eur. Phys. J. C \textbf{76}, 521 (2016)},
\href{https://arxiv.org/abs/1606.05452}{\color{teal}{[arXiv:1606.05452 [hep-ph]]}}.


\bibitem{Najjar:2025dzl}
Z.~R.~Najjar and K.~Azizi,
Investigation of triply heavy spin-3/2 baryons in their ground and excited states,
\href{https://doi.org//10.1016/j.physletb.2025.140000}{Phys. Lett. B \textbf{871}, 140000 (2025)},
\href{https://arxiv.org/abs/2504.06822}{\color{teal}{[arXiv:2504.06822 [hep-ph]]}}.

\bibitem{Ioffe:1981kw}
B.~L.~Ioffe,
Calculation of Baryon Masses in Quantum Chromodynamics,
\href{https://doi.org/10.1016/0550-3213(81)90259-5}{Nucl. Phys. B \textbf{188}, 317-341 (1981)}.

\bibitem{Aliev:2002ra}
T.~M.~Aliev, A.~Ozpineci and M.~Savci,
Octet baryon magnetic moments in light cone QCD sum rules,
\href{https://doi.org/10.1103/PhysRevD.67.039901}{Phys. Rev. D \textbf{66}, 016002 (2002)},
\href{https://arxiv.org/abs/hep-ph/0204035}{\color{teal}{[arXiv:hep-ph/0204035 [hep-ph]]}}.

\bibitem{Azizi:2014yea}
K.~Azizi and N.~Er,
Properties of nucleon in nuclear matter: once more,
\href{https://doi.org/10.1140/epjc/s10052-014-2904-5}{Eur. Phys. J. C \textbf{74}, 2904 (2014)},
\href{https://arxiv.org/abs/1401.1680}{\color{teal}{[arXiv:1401.1680 [hep-ph]]}}.

\bibitem{Ozdem:2017jqh}
U.~Ozdem and K.~Azizi,
Magnetic and quadrupole moments of the $Z_c(3900)$,
\href{https://doi.org//10.1103/PhysRevD.96.074030}{Phys. Rev. D \textbf{96}, 074030 (2017)},
\href{https://arxiv.org/abs/1707.09612}{\color{teal}{[arXiv:1707.09612 [hep-ph]]}}.

\bibitem{Azizi:2018duk}
K.~Azizi, A.~R.~Olamaei and S.~Rostami,
Beautiful mathematics for beauty-full and other multi-heavy hadronic systems,
\href{https://doi.org//10.1140/epja/i2018-12595-1}{Eur. Phys. J. A \textbf{54}, 162 (2018)},
\href{https://arxiv.org/abs/1801.06789}{\color{teal}{[arXiv:1801.06789 [hep-ph]]}}.



\bibitem{Azizi:2017ubq}
K.~Azizi and N.~Er,
X (3872): propagating in a dense medium,
\href{https://doi.org//	10.1016/j.nuclphysb.2018.09.014}{	Nucl. Phys. B \textbf{936}, 151-168 (2018)},
\href{https://arxiv.org/abs/1710.02806v3}{\color{teal}{	[arXiv:1710.02806 [hep-ph]]}}.

\bibitem{Belyaev:1982sa}
V.~M.~Belyaev and B.~L.~Ioffe,
Determination of Baryon and Baryonic Resonance Masses from QCD Sum Rules. 1. Nonstrange Baryons,
\href{https://inis.iaea.org/records/r9jqb-2k079}{Sov. Phys. JETP \textbf{56}, 493-501 (1982)}.

\bibitem{DELPHI:1993ukk}
P.~Abreu \textit{et al.} [DELPHI],
Determination of alpha-s from the scaling violation in the fragmentation functions in $e^+ e^-$ annihilation,
\href{https://doi.org//10.1016/0370-2693(93)90587-8}{Phys. Lett. B \textbf{311}, 408-424 (1993)}.

\bibitem{Belyaev:1982cd}
V.~M.~Belyaev and B.~L.~Ioffe,
Determination of the baryon mass and baryon resonances from the quantum-chromodynamics sum rule. Strange baryons,
\href{http://jetp.ras.ru/cgi-bin/dn/e_057_04_0716.pdf}{Sov. Phys. JETP \textbf{57}, 716-721 (1983)}.

\bibitem{Pospelov:2005pr}
M.~Pospelov and A.~Ritz,
Electric dipole moments as probes of new physics,
\href{https://doi.org//10.1016/j.aop.2005.04.002}{Annals Phys. \textbf{318}, 119-169 (2005)},
\href{https://arxiv.org/abs/hep-ph/0504231}{\color{teal}{[arXiv:hep-ph/0504231 [hep-ph]]}}.

\bibitem{Wang:2018kto}
Q.~W.~Wang, S.~X.~Qin, C.~D.~Roberts and S.~M.~Schmidt,
Proton tensor charges from a Poincar{\'e}-covariant Faddeev equation,
\href{https://doi.org//10.1103/PhysRevD.98.054019}{Phys. Rev. D \textbf{98}, 054019 (2018)},
\href{https://arxiv.org/abs/1806.01287}{\color{teal}{[arXiv:1806.01287 [nucl-th]]}}.

\bibitem{Agaev:2020zad}
S.~Agaev, K.~Azizi and H.~Sundu,
Four-quark exotic mesons,
\href{https://doi.org//10.3906/fiz-2003-15}{Turk. J. Phys. \textbf{44}, 95 (2020)},
\href{https://arxiv.org/abs/2004.12079v1}{\color{teal}{[arXiv:2004.12079 [hep-ph]]}}.

\end{thebibliography}
\end{document}